\documentclass{article}

\usepackage{setspace}
\usepackage{geometry}
\usepackage[utf8]{inputenc}
\usepackage{amsmath}
\usepackage{amssymb}
\usepackage{amsthm}
\usepackage{graphicx}
\usepackage{natbib}
\usepackage{caption}
\usepackage{subcaption}
\usepackage{float}

\newtheorem{theorem}{Theorem} 

\newtheorem{lemma}[theorem]{Lemma}

\DeclareMathOperator*{\argmin}{arg\,min}

\title{A simultaneous confidence bounded true discovery proportion perspective on localizing differences in smooth terms in regression models}
\author{David Swanson\footnote{email: dmswanson@mdanderson.org, ORCID: 0000-0003-3174-1656} \\ \small{{\it University of Texas MD Anderson Cancer Center, Department of Biostatistics, Houston, TX}}}
\date{} 

\begin{document}

\maketitle

\begin{abstract}
A method is demonstrated for localizing where two spline terms, or smooths, differ using a true discovery proportion (TDP) based interpretation. 
The procedure yields a statement on the proportion of some region where true differences exist between two smooths, which results from use of hypothesis tests on collections of basis coefficients parameterizing the smooths.  The methodology avoids otherwise ad hoc means of making such statements like subsetting the data and then performing hypothesis tests on the truncated spline terms.  TDP estimates are 1-$\alpha$ confidence bounded simultaneously.  This means that the TDP estimate for a region is a lower bound on the proportion of actual difference, or true discoveries, in that region with high confidence regardless of the number of regions at which TDP is estimated.  Our procedure is based on closed-testing using Simes local test.  This local test requires that the `multivariate $\chi^2$ test statistics of generalized Wishart type' underlying the method are positive regression dependent on subsets (PRDS), which we show.  The method is well-powered because of a result on the off-diagonal decay structure of the covariance matrix of penalized B-splines of degree two or fewer.  We show consistency of the procedure, demonstrate achievement of estimated TDP in simulation and analyze a study of walking gait of cerebral palsy patients.

\vspace{0.1cm}
\end{abstract}

\section{Introduction}

Methods to model non-linear curves have manifold applications in statistics.  Many regression modelling problems necessitate more flexible assumptions than linearity, and approximating non-linearities with a collection of spline basis functions is common \citep{wahba_spline_1990,hastie_generalized_1990,eilers_flexible_1996,reiss_smoothing_2009}.  The variety of modelling techniques with splines and means of testing hypotheses on them has increased to where an analyst now has a wide variety of methods to perform desired analyses.

Hypothesis testing on splines is an area that has benefited from recent development.  Since the study of \citet{crainiceanu_exact_2005} into exact testing for linearity on a single smooth, a term we use interchangeably with spline, there has been fruitful work on testing different hypotheses on smooths, such as their linearity or whether a collection of smooths are in fact different from one another.  
Permutation, Bayesian, and frequentist approaches and interpretations have all been proposed \citep{crainiceanu_likelihood_2004,fitzmaurice_note_2007,wood_p-values_2013,nychka_bayesian_1988}.  

While hypothesis testing methodology on entire smooths is in a mature state, a relevant yet much less developed goal is identifying specific regions where two or more curves differ.  Oftentimes in analyses identifying regions of difference and making statistical statements on them is important for inference, for example when trying to localize where walking gait differs for cerebral palsy patients using different prosthetics, an application we describe later \citep{skaaret_comparison_2019,skaaret_impact_2019,roislien_simultaneous_2009}.

Broadly, our proposal for localizing differences between smooths relies on testing collections of estimated spline coefficients, which correspond to hypothesis tests of equivalence of two strata in regions defined by the coefficients when an appropriate basis set is chosen.  Determining regions on which the two strata differ becomes a problem of rejecting groups of these hypotheses.

However, difficulties remain because of the way basis functions can influence the entire smooth, high degree of correlation of estimated parameters, and shrinkage of otherwise overly flexible spline bases.  Other challenges with multiple testing arise if the analyst wants to perform many exploratory analyses, such as testing the equivalence of two strata in several different regions while controlling type 1 error.  

We address these challenges by using regularized B-spline bases, or p-splines, and adapting a testing procedure described in \citet{hommel_stagewise_1988} that allows for circular testing -- that is, results of hypothesis tests can motivate new ones without inflating type 1 error rates.  The procedure would be computationally intractable if not for building on Simes local tests, the use of which has been made tractable in closed testing in \citet{goeman_simultaneous_2019} and which we further develop for use in our unique regression setting. 
The method yields a true discovery proportion (TDP) based interpretation, which 
in this context is a statement on the proportion of some region where true differences exist between two smooths.  
The granularity of specifying regions is a function of the knots for the smooth terms, which can be placed at fine grain because parameter penalization is used.  TDP statements are made with 1-$\alpha$ confidence that the TDP estimate for a region is a lower bound on the actual proportion of true discoveries in that region simultaneously.  The simultaneity of the confidence bound allows one to compare two strata on as many regions as one wants without inflating type 1 error, in this case error interpreted as an overestimate of TDP on a region.



In Section \ref{sec:methods}, we describe the procedure, its basis in closed-testing, and show that the positive regression dependence on subsets (PRDS) condition, sufficient for Simes inequality, holds on the multivariate $\chi^2$ test statistics used in the procedure. 
In Section \ref{sec:results}, we vary the degree of differences between two smooths and prescribed $\alpha$ and show their influence on estimated TDP.  In Section \ref{data_analy}, we apply the procedure to a study of walking gait in cerebral palsy patients to localize where gait differences occur between strata of two different prosthetic types. 
We conclude in Section \ref{sec:discussion} with a discussion of extensions to arbitrary spline bases, more than two smooths, and small sample sizes.







\section{Methods}
\label{sec:methods}

\subsection{Background}


Suppose we have two independent strata of outcomes and covariates comprising our data, $\{ {\pmb y_{l}},  X_{l}, {\pmb z_{l}}\}$, ${\pmb y_{l}}=(y_{l1}, \cdots ,  y_{l,n_l})^T$ a length $n_l$ vector of outcomes, $X_{l} = \left[{\pmb x_{l1}} , \cdots ,{\pmb x_{l,n_l}}\right]^T$, a $n_l \times p$ matrix of covariates, ${\pmb z_{l}}=(z_{l1}, \cdots ,  z_{l,n_l})^T$ a covariate vector to be modelled with a smooth, $l\in 1,2$, for respective sample sizes $n_1$ and $n_2$, and where ${\pmb s}^T$ is the transpose of its argument ${\pmb s}$. Consider for the moment only one stratum $l$.
When we model the $i^{th}$ outcome $y_{li}$, $i \in \{1,2,\dots , n_l\}$, assumed exponential family distributed, with a smooth function of some covariate $z_{li}$ and conditioning on a $p$-vector ${\pmb x_{li}}$ treated as a fixed effect, we often do so with 
$$g(u_{li}) = {\pmb x}_{li}^T {\pmb \beta}_l + \sum_{k=1}^m b_{lk} \, f_{k}(z_{li})$$ where $u_{li} = E(y_{li} | {\pmb x_{li}}, z_{li})$ for expectation $E(\cdot)$, spline basis functions $f_{k}(\cdot)$, and link function $g(\cdot)$ \citep{wahba_spline_1990}.  We use ${\pmb b_l}= (b_{l1}, \cdots , b_{lm})^T$, ${\pmb \beta_l}= (\beta_{l1} ,  \cdots , \beta_{lp})^T$, and call the associated log-likelihood for all $n_l$ samples $\, {\pmb \ell}(\pmb{b_l},{\pmb \beta_l}, \phi_l | {\pmb y_l})$, where $\phi_l$ is a dispersion parameter.  Our goal is identifying where the last term in the model, call it $h_l(z) = \sum_{k=1}^m b_{lk} \, f_k(z)$, a smooth function composed of basis functions $\{f_k \}$, differs between strata with respect to $z$.  



A wide variety of basis expansions have been described to model smooth terms such as $h_l(z)$, and here we develop our procedure with B-splines \citep{de_boor_practical_1978}.  This expansion spans the same large space of smooth functions as cubic splines when third degree basis functions are used, for example, and so has continuous second derivatives.  However, these basis functions are additionally only non-zero on a compact region which can be small relative to the support of the respective covariate
if one chooses a sufficient number of knots \citep{ruppert_selecting_2002}.  The linear predictor with respect to any $z_{li}$ is therefore a function of only $d+1$ parameters scaling the B-spline basis set for $d$ the degree of the basis functions and will be chosen as 2 or 3 in most applications, as we examine and use here.  This stands in contrast to other spline bases where certain fitted values may be a function of many basis functions. 

Generally shrinkage methods are necessary to constrain the oftentimes high-dimension ${\pmb b_l}$ parameter \citep{craven_smoothing_1979,golub_generalized_1979,wood_smoothing_2016}.  In this case one augments the objective function ${\pmb \ell_l}$ to be $$ {\pmb \ell}(\pmb{b_l},{\pmb \beta_l}, \phi_l | {\pmb y_l}) - \lambda_l \,{\pmb b_l}^T S \, {\pmb b_l}\, /\, 2 $$
for some penalization matrix $S$ and scaling parameter $\lambda_l$ which is often estimated with restricted maximum likelihood (REML) or generalized cross validation (GCV) \citep{wahba_spline_1990}. 
When penalizing B-splines, which are termed P-splines, we use a penalty matrix composed of second order differences on adjacent ${b_{lk}}$'s.  The difference matrix $D$ is then $(m-2) \times m$, with 3 non-zero elements, 1, -2, 1 along each row and aligned with the diagonal, and then we construct $S = D^T D$. 

\subsection{Testing differences of knot-defined intervals between strata}

 If we estimate $h_l(z) = \sum_{k=1}^m b_{lk} \, f_k(z) $ for each $l$ on a common set of knots, ${\pmb \kappa} = (\kappa_1 , \kappa_2, \dots , \kappa_{m-d+1})$, where the endpoints encompass extrema of the covariate $z$'s support and $m$ is the dimension of the B-spline basis, we can localize where $h_l(z)$ differs between strata by formulating the question in terms of hypothesis tests on the $b_{lk}$'s.  So define $\Delta {\pmb b} = ({\pmb b_1} - {\pmb b_2})$ and the $\Delta b_k$'s as its composing elements.  For an $m$-dimensional ${\pmb b_l}$ there are $m_T = m-d$ intervals along the covariate's support defined by the knots ${\pmb \kappa}$.  Then the region defined ${\mathcal R_k} = (\kappa_{k}, \kappa_{k+1})$ for $1\leq k \leq m_T$ is entirely determined by the $d+1$ scalars $b_{l,k}, b_{l,k+1} \dots, b_{l,k+d}$ for the $l^{th}$ stratum.  Call $$H_k^0: (\Delta b_k ,\dots , \Delta b_{k+d})={\pmb 0} \mbox{ for  } 1\leq k \leq m_T$$ 
the hypothesis of equivalence of the two stratum-specific smooths on this region ${\mathcal R_k}$, which is one of the elementary hypotheses of our procedure upon which local tests and then TDP estimation are built. Thus, with estimates ${\hat b_{l,k}}$ for all $l,k$ and their covariances, one can use a normal approximation to test $H_k^0$ for any exponential family model.  When the sample size is small and ${\pmb y_l}$ is conditional Gaussian, one can use Hotelling's T$^2$ for these tests \citep{hotelling_generalization_1931}, but otherwise a $\chi^2$ test performs adequately as simulations show.  
 
 To consider estimation, assume for exposition that the elements $(y_{l1} , \cdots , y_{l,n_l})^T =  {\pmb y_l}$ are independent Gaussian conditional on $(X_l, {\pmb z_l})$.  Then for fixed $\lambda_l$ the penalized least squares estimates for $({\pmb \beta_l}, {\pmb b_l})$  
 with $l \in \{1,2 \}$ are

\begin{equation}
( {\pmb {\hat \beta}_l}, {\pmb {\hat  b}_l})^T = \argmin_{{\pmb \beta_l},{\pmb b_l}}\sum_{i=1}^{n_l} \biggl(y_{li} - \sum_{j=1}^p x_{lij} \beta_{lj} -  h_l(z_{li}) \biggr)^2 + \lambda_l \, {\pmb b_l}^T S \, {\pmb b_l} 
\label{objec_fcn}
\end{equation}
where $S$ is not indexed by $l$ because ${\pmb \kappa}$ is invariant to stratum.
The $({\pmb {\hat \beta_l}}, {\pmb {\hat b_l}})$ satisfy 
\[ \begin{bmatrix}
  {\pmb {\hat \beta}_l}  \\
  {\pmb {\hat b}_l}
\end{bmatrix} = 
C_l^{-1}  \begin{bmatrix}
  X_l^T {\pmb y_l} \\
  Z_l^T {\pmb y_l}
\end{bmatrix} \]
where 
\[ 
C_l = \begin{bmatrix}
  X_l^T X_l  & X_l^T Z_l  \\
  Z_l^T X_l  &  Z_l^T Z_l + \lambda_l S
\end{bmatrix}
\]
and where $X_l$ is the $n_l \times p$ fixed effects design matrix for stratum $l$ and $Z_l$ is the $n_l \times m$ spline basis expansion for covariate vector ${\pmb z_l}$ with entry $f_k(z_{li})$ at the $i, k$ element of stratum $l$.

To estimate the covariance of the ${\pmb {\hat b_l}}$'s in the Gaussian case, consider the objective function in (\ref{objec_fcn}).  For fixed and known $\lambda_l$, one can take a Bayesian or frequentist perspective on the covariance of ${\pmb b_l}$ or ${\pmb {\hat b}_l}$, respectively \citep{wood_confidence_2006,marra_coverage_2012,nychka_bayesian_1988,wahba_bayesian_1983}.  Under the Bayesian framework, one can consider the generalized inverse of rank-deficient $\lambda_l  S$ as a prior covariance on ${\pmb b_l}$ so that $${\pmb b_l} \sim MVN({\bf 0}, \lambda_l \, S^{-})$$ where $S^{-}$ is the Moore-Penrose generalized inverse.  Such an approach ignores the variability of choosing $\lambda_l$, but doing so is of less influence for the estimated covariance and one can use a correction if desired \citep{wood_smoothing_2016}.



Under this Bayesian conceptualization and following \cite{wahba_bayesian_1983} and \cite{nychka_bayesian_1988}, the posterior covariance of ${\pmb b_{l}}$ when $X_l$ contains no covariates is 
\begin{equation}
V_l = {\hat \phi_l}\,(Z_l^{T} Z_l + \lambda_l \, S)^{-1}
\label{covar}
\end{equation}
We can consider this case without $X_l$ in part because $Z_l$ spans the intercept which is left unpenalized by $S$.  When $X_l$ is nonempty, $V_l$ is the inverse of the Schur complement of the upper $p\times p$ block of $C_l$.
We use expression (\ref{covar}) for the covariance because it will tend to be slightly conservative compared to that from the frequentist perspective, which is additionally multiplied by $Z_l^T Z_l (Z_l^T Z_l + \lambda_l \, S)^{-1}$, a term with determinant less than 1 for positive $\lambda_l$.

With estimates of ${\pmb {\hat b}_l}$ and their variances, we can test elementary hypothesis $H_k^0$ for $(b_{1,k},\dots ,  b_{1,k+d}) = (b_{2,k},\dots ,  b_{2,k+d})$, equivalence of smooths on region ${\mathcal R_k}$, for each $k$. We construct the associated test statistic $T_k$ with 
\begin{equation}
T_k = ({\pmb {\hat b}_1} - {\pmb {\hat b}_2})_{k,d}^T \; ([V_1]_{k,d} + [V_2]_{k,d})^{-1} \; ({\pmb {\hat b}_1} - {\pmb {\hat b}_2})_{k,d},
\label{quad_eqn}
\end{equation}
where $[V_l]_{k,d}$ is the submatrix along the diagonal of $V_l$, starting at row and column index $k$ and extending to $k+d$, and $(v)_{k,d}$ is a subvector of vector $v$ corresponding to indices $k, \dots , k+d$. 
Because ${\pmb {\hat b}_1}$ and ${\pmb {\hat b}_2}$ are fit on different strata of data and therefore independent, their covariances are additive.  Parameterizing the stratum-specific models into a single model with strata-smooth interactions poses no gain in statistical efficiency since the flexibility needed to perform the desired inference renders fitted values and log-likelihoods identical to the approach with stratum-specific models.  The proposed collection of test statistics $\{T_k \}$ can be shown to exhibit the following asymptotic behavior: 
\begin{lemma}
$T_k$ is asymptotically distributed $\chi_{d+1}^2$ under $H_k^0$, that is $({\pmb b_1} - {\pmb b_2})_{k,d} ={\pmb 0}$, choosing the tuning parameter by REML or GCV and using either the frequentist or Bayesian covariance estimator. 
\label{lem:consistency}
\end{lemma}
\noindent
We show Lemma \ref{lem:consistency} in the Supplementary Material, which depends on rotation of the covariance matrix and the influence of penalization vanishing asymptotically.  By comparing $T_k$ against $\chi_{d+1}^2$ one obtains p-values $p_k$ distributed uniform under the null hypothesis.  There are $m_T$ such test statistics and p-values, one for each hypothesis of region-wise ${\mathcal R_k}$ equivalence of the two strata. 


Calculating the test statistics according to (\ref{quad_eqn}) requires that $[V_1]_{k,d} + [V_2]_{k,d}$ be inverted along a sliding window of sets of parameters for each $k\in \{1, \dots , m_T\}$.  Since adjacent submatrices overlap except on one index, one can reuse information from one inversion to simplify the next.  By performing an appropriate rank 1 deletion and addition to generate each subsequent inverse, one can reduce the manageable computational burden from $O(m_T(d+1)^3)$ to $O(2m_T(d+1)^2 + (d+1)^3)$, which is a meaningful reduction for small $d$ when $m_T$ is large as in our setting.

\subsection{True discovery proportion estimation and closed testing}

With p-value $p_k$ for each elementary hypothesis $H_k^0$ of strata equivalence on ${\mathcal R_k}$, we seek to build a procedure for estimating a simultaneous confidence bounded true discovery proportion on arbitrary regions ${\mathcal R}_R = \cup_{k\in R} {\mathcal R}_k$ for an index set $R\subseteq \{1,\dots , m_T \}$.  Call $H_R = \cap_{k\in R} H_k^0$ the intersection hypothesis associated with equivalence of strata on region ${\mathcal R}_R$.  We propose estimating the true discovery proportion for $H_R$ using an approach grounded in closed testing.  

\citet{hommel_multiple_1986} describes a closed-testing procedure whereby if for each $H_R$ there exists a ``local'' test of size $\alpha$ and if the local test rejects $H_J$ for every $R\subseteq J$, $H_R$ is rejected under the closed-testing regime.  This procedure gives strong control of the family wise error rate (FWER) at level $\alpha$ for all intersection hypotheses simultaneously.  Define $${\mathcal X} = \{ R : H_R\mbox{ is rejected by the closed testing procedure}\}$$
Described this way, one sees that identifying elements of ${\mathcal X}$ requires enumeration of the $2^{m_T}$ power set of intersection hypotheses, making computation intractable for even small $m_T$ were one to apply the test statistic in (\ref{quad_eqn}) or most others as the local test. However, shortcuts have been proposed for certain local tests.  \citet{hommel_stagewise_1988} provided a procedure to identify elementary hypotheses belonging to ${\mathcal X}$ for Simes local tests, which \citet{goeman_multiple_2011} and \citet{goeman_simultaneous_2019} extended to intersection hypotheses for this same local test. Simes test is described in \citep{simes_improved_1986}, who showed that for uniformly distributed p-values which are either independent or fall into certain families of dependence structures, one can use
\begin{equation}
p_{(k)} \leq k \alpha/m_T \;\;\;\mbox{ for at least one }k\in \{1,\dots , m_T \}
\label{simes}
\end{equation}
as a rejection region for a test which has type 1 error bounded by $\alpha$, where $p_{(k)}$ is the $k^{th}$ ordered p-value, and where $\alpha$ is achieved when the $p_k$ are independent. The inequality forms the basis for the eponymous test, which rejects when (\ref{simes}) holds.  
\citet{benjamini_control_2001} showed that the positive regression dependence on subsets (PRDS) condition for the p-values $\{p_k\}$ is sufficient for the validity of the test. Application of a monotone transformation to all $\{p_k\}$, including the map back to $\{T_k\}$, maintains PRDS.  We show the condition holds for the $T_k$'s in the next section. 

\citet{goeman_multiple_2011} point out the connection between membership in ${\mathcal X}$ for intersection hypotheses $H_R$ and estimation of the false discovery proportion, the complement of which is TDP.  In particular, the size of the largest index set $S$ of $R$ such that $S \notin {\mathcal X}$ is the confidence bounded false discoveries of $R$.   So for intersection hypothesis $H_R$ on index set $R$, the simultaneous $1-\alpha$ lower confidence bound for the number of true discoveries in the set is $$\phi_{\alpha}(R)= \#R - \max \{\#S : S\subset R, S \notin {\mathcal X} \}, $$
normalization of which yields TDP and where $\#(\cdot)$ is the cardinality of its argument. When using Simes local tests, the quantity admits a shortcut with
\begin{equation}
\phi_{\alpha}(R)= \max_{1\leq u \leq \# R} 1 - u + \#\{i \in R : h_\alpha \, p_i \leq u\, \alpha \}
\label{phi_eqn}
\end{equation}
where
\begin{align}
h_\alpha = \max \Bigl\{0\leq i \leq m_T : \{p_{(m_T-i+1)},p_{(m_T-i+2)},\dots , p_{(m_T)}\}\mbox{ is not rejected by Simes test }\Bigr\} 
\label{h_eqn}
\end{align}
so $h_\alpha$ need only be calculated once for all different $R$ for which TDP is to be estimated.  We then use $\phi_{\alpha}(R)/\#R$ as the simultaneous $1-\alpha$ lower confidence bound for the TDP of region ${\mathcal R}_R$ corresponding to the intersection hypothesis $H_R$, where $\alpha$ is that level used for the local test, the interpretation of which is the typical one in hypothesis testing: a level $\alpha$ test will reject under the null with proportion $\alpha$.  Whereas, the interpretation of the 1-$\alpha$ simultaneous confidence bound for TDP is that the bound will overestimate the actual TDP with proportion $\alpha$.  Because that bound is simultaneous, it holds for as many applications of the procedure to regions ${\mathcal R}_R$ as desired.  

So the validity of Simes test relies on $\{T_k\}$ being PRDS.  And the shortcuts necessary for feasibility of the closed-testing procedure rely on use of Simes test. This procedure can then be summarized with the following algorithm: 

\begin{enumerate}

\item[{\bf Step 1}] Fit stratum-specific models on a common set of knots such that one obtains estimates of $b_{lk}$ and covariance matrices $V_l$ for $l=1,2$ and $k\in \{1,\dots , m\}$.

\item[{\bf Step 2}] For each region, ${\mathcal R}_k$, calculate a p-value $p_k$ for hypothesis $H_k^0:(b_{1,k},\dots ,  b_{1,k+d}) = (b_{2,k},\dots ,  b_{2,k+d})$, equivalence understood element-wise, using equation (\ref{quad_eqn}).

\item[{\bf Step 3}] Calculate $h_\alpha$ once for the set of all $p_k$, $1\leq k \leq m_T$, with equation (\ref{h_eqn}).

\item[{\bf Step 4}] For every ${\mathcal R}_R$ of interest, find $u$ in equation (\ref{phi_eqn}), thus estimating the confidence bounded TDP $\phi_\alpha(R)$ for index set $R$.

\end{enumerate}
Now we examine the validity of PRDS on the collection of $T_k$.







\subsection{PRDS on multivariate $\chi^2$ random variables}
\label{sec:simes}


A set of test statistics $\{T_1,\dots , T_{m_T}\}$ is PRDS on the subset $I=\{1\dots m_T \}$ if $$P( \pmb{T}_{\backslash k } \succ \pmb{t}_{\backslash k}  | T_k=t_k ) $$
is non-decreasing in $t_k$ for any $\pmb{t}_{\backslash k}\in \mathbb{R}^{m_T-1}$ and all $k$, where a vector $\pmb{s}_{\backslash k}$ is understood $(s_1,\dots , s_{k-1},s_{k+1},\dots , s_{m_T} )$ and $\succ$ is taken element-wise (cf. \citet{lehmann_concepts_1966}). \citet{benjamini_control_2001} showed that this condition is sufficient for the validity of the test expressed in (\ref{simes}), assuming the corresponding p-values $\{p_k\}$ are distributed uniform under the null hypothesis.  In contrast to other settings in which PRDS is unexamined but sometimes implicitly assumed, such as genome-wide association studies (GWAS) or medical image analysis \citep{rosenblatt_all-resolutions_2018}, for splines we estimate all parameters in a single regression model.  This renders many of the underlying ${\hat b}_{l,k}$'s negatively correlated because of positive correlation in the information matrix due to overlapping B-spline basis functions.  This stands in contrast to image analysis or GWAS because test statistics in those settings are often calculated from many univariate regressions.  The covariance matrix of these test statistics is then the correlation of the design matrix itself.  That design matrix will tend to be, at least locally, positively correlated.
Care must be taken in showing this condition in our unique regression setting, which is additionally complicated by the quadratic form structure of test statistics.  Adjacent test statistics also share elements of their covariance matrices because they are taken from a sliding window along the diagonal of the covariance matrix of basis coefficients. 

There are specific cases in which PRDS is proven to hold and includes multivariate Gaussian test statistics whose inverse covariance is an M-matrix \citep{karlin_total_1981}.  A non-singular covariance matrix is M if all entries are non-negative, and all off-diagonal entries of the inverse are non-positive.  The condition covers the case of bivariate Gaussian, positively correlated test statistics for example.  Toeplitz matrices must satisfy a specific decay of off-diagonal entries for their inverses to be M, which will not generally hold for the multidiagonal type Toeplitz structure under consideration because of the sharp decline to 0 at some off-diagonal entry.  Nevertheless, the validity of  (\ref{simes}) is believed to approximately hold in most practical cases of positive correlation \citep{finner_simes_2017,sarkar_probability_1998}.






\noindent
{\bf Definition 1.} {\it Consider random variables $(U_{1,1}\dots U_{1,v_1},U_{2,1} \dots  U_{2,v_2} \dots \dots  U_{m,1} \dots U_{m.v_m})$, each standard normal with $\mbox{cor}(U_{ij},U_{kl})=\rho_{ij,kl}$ where $\rho_{ij,kl}=0$ if $i=k$.  Construct $Q_i = \sum_{j=1}^{v_i} U_{ij}^2 $.  Then ${\pmb Q} =(Q_1, \dots , Q_m)$ is multivariate $\chi^2$ of generalized Wishart type on parameters ${\pmb Q}\sim \chi^2 (m, {\pmb v}, W)$, with ${\pmb v}=(v_1, \dots , v_m)$ and $W$ parameterizing the correlation between the $U_{ij}$'s} \citep{dickhaus_simultaneous_2016}. 

One can see that the test statistics $\{T_k\}$ are multivariate $\chi^2$, having arisen from quadratic forms of test statistics normalized by their own covariance. Because of the sliding window used in calculating $\{T_k\}$ in our application, $W$ can have elements of 1 among the non-zero $\rho_{ij,kl}$.  Regardless, here we claim that these distributions are PRDS.  
\begin{theorem}
The multivariate $\chi^2$ distribution of generalized Wishart type ${\pmb Q}\sim \chi^2 (m, {\pmb v}, W)$ is PRDS with the subset $i\in \{1, \dots , m\}$.  
\end{theorem}

\noindent
The proof is given in the Supplementary Material.  Briefly, one wants to show $P(Q_1,\dots ,Q_{i-1},Q_{i+1}, \dots Q_m | Q_i=r)$ is stochastically increasing in each element as a function of $r$. One can consider $Q_i$ as a mixture distribution arising from $(U_{i,1}\dots U_{i,v_i})$, the domain of which is a $v_i$-sphere of radius say $\sqrt{r_1}$ because $Q_i = \sum_{j=1}^{v_i} U_{ij}^2$ by definition and so $\sum_{j=1}^{v_i} U_{ij}^2=r_1$ by the conditioning event, values consistent with which describe the $v_i$-sphere.  The density of $(U_{i,1}\dots U_{i,v_i})$ is a function of the intersection of that hypersphere with the multivariate Gaussian with identity covariance from which the $U_{ij}$'s are drawn. Consider an arbitrary $Q_k$ with $k\neq i$.  For a fixed $(U_{i,1}\dots U_{i,v_i})$ and using properties of the conditional multivariate Gaussian, it is not difficult to calculate the conditional density of  $(U_{k,1}, \dots ,U_{k,v_k})$, from which $Q_k$ is calculated with $\sum_{j=1}^{v_k} U_{kj}^2$ .  For this fixed $(U_{i,1}\dots U_{i,v_i})$, then, and using a result from \citet{imhof_computing_1961}, $Q_k$ will be distributed as a non-central, scaled sum of $v_k$ $\chi^2_1$'s, which will be one infinitessimal element of its mixture.  
Within this framework, one can define a bijection
mapping $(U_{i,1}\dots U_{i,v_i})$ on the $v_i$-sphere with radius $\sqrt{r_1}$ to the $v_i$-sphere with radius $\sqrt{r_2}$ with $\sqrt{r_2}>\sqrt{r_1}$ and examine how the distribution of $Q_k$ changes. 
In doing so, one can directly compare distributions composing the mixtures, whose parameters and mixing probabilities will be the same with the exception of increasing non-centrality parameters, giving the result. 

\subsection{Statistical power of the procedure}
\label{sec:power}

We argue that $\alpha$ is nearly attained in expression (\ref{simes}) by demonstrating that p-values are practically independent for all but relatively proximal ones with respect to the region being tested.  This characteristic justifies the well-poweredness of the method, 
which we show by giving an analytic inverse of multidiagonal Toeplitz matrices with modification to the corner elements, which is the form of the sum of the information and penalization matrices when the covariate is uniformly distributed.

The inverse of the multidiagonal Toeplitz matrix is proportional to the covariance of $(T_1 \dots T_{m_T})'$, and off-diagonal elements of this inverse are shown to have log-linear decay and in practice approach 0 quickly. 
Small deviations from the Toeplitz structure can be addressed with the approximation $(A + \varepsilon Q)^{-1} \approx A^{-1} - \varepsilon A^{-1} Q A^{-1}$ 
for small $\varepsilon$.  The information matrix is multidiagonal regardless of the covariate's distribution, with all entries 0 except the first $d$ off the diagonal for splines of degree $d$.  That is, second degree splines result in a pentadiagonal matrix, where a total of $5$ diagonals along the matrix are non-zero, centered at the main diagonal.  B-spline expansions have this structure because the support of a single basis function is compact relative to the covariate's support.

We show log-linear decay in the covariance of proximal $b_{l,k}$'s, within strata of $l$, for up to second degree splines, by giving an analytic result for the penalized covariance matrix in equation (\ref{covar}).  We do so by showing a factorization of (\ref{covar}) into two tridiagonal matrices, which we invert and analyze.  If the splines are first degree, then the tridiagonal inverse suffices.  \citet{dow_explicit_2003} gives results on linear difference equations necessary to generalize to higher order splines, while \citet{bickel_approximating_2012} provide a log-linear bound on off-diagonal elements of inverses of banded positive semidefinite matrices, applicable for spline expansions of any degree when the information matrix dominates the penalty.  Proofs for the following are found in the Supplementary Material.



\begin{lemma}
Consider the pentadiagonal Toeplitz matrix with modified corner elements

\[ P = 
 \begin{bmatrix}
  \epsilon-\zeta_1  &  \theta & \lambda & 0 & 0 & \cdots & \cdots & 0 \\
    \theta  &  \epsilon & \theta & \lambda & 0 & 0 & \cdots & 0 \\
  \lambda  &  \theta & \epsilon & \theta & \lambda & 0 & \cdots & 0 \\
 0 & \lambda  &  \theta & \epsilon & \theta & \lambda &  \cdots & 0 \\  
 \vdots & \vdots  &  \ddots & \ddots & \ddots & \ddots &  \vdots & \vdots \\
 0 & \cdots  &  0 & 0 & \lambda & \theta &  \epsilon & \theta \\
 0 & \cdots & \cdots  &  0 & 0 & \lambda & \theta &  \epsilon -\zeta_2 \\ 
\end{bmatrix} \]
Then provided $\theta^2 - 4\lambda (\epsilon - 2\lambda) > 0$, there exists a factorization of $P$ into two real tridiagonal toeplitz matrices, $Z_1$ and $Z_2$ with diagonal elements $(\lambda,\pi_1,\lambda)$ and $(1,\pi_2/\lambda,1)$, respectively, where $\pi_1,\pi_2$ are roots of the polynomial in $s$, $s^2 - \lambda s + \lambda(\epsilon - 2\lambda)$, and where we then have $\zeta_1, \zeta_2 = \lambda$.  
\label{factor}
\end{lemma}

\begin{theorem}
Elements of $P^{-1}$ exhibit log-linear decay in absolute value along its off-diagonal at rate $\min_i \{\psi_i\}$ for $\psi_i = \operatorname{arcosh} (\pi_i/(2\lambda))>0$ if $\pi_i^2 > 4 \lambda^2$ for $i=1,2$
\label{log_linear}
\end{theorem}

Though the penalization and information matrices both satisfy conditions of Lemma \ref{factor}, their sum may not depending on the penalty parameter because of negative elements in the penalization matrix.  One must assume the information matrix dominates the penalty, which will hold asymptotically and in most practical settings including our data analysis.  Since our procedure is most appropriate with many knots, a moderate to large sample size is therefore encouraged from different aspects of the methodology.  If the condition were ever to not hold on the sum of these matrices, one could simply reduce the penalization parameter, thereby slightly increasing variance in parameter estimation while reducing bias.

Assuming a uniformly-distributed covariate, the covariance matrix varies with the shrinkage parameter scaling the penalization matrix.  Also, for the case of second degree B-splines, one must slightly translate two of the outermost knots so that the corner element structure from Theorem \ref{log_linear} holds.  If one assumes the information dominates the penalization matrix, there may be little variability in the covariance matrix and therefore also in $\min_i \psi_i$ across analyses. In practice one may therefore assert that generally the covariance matrix approaches $0$ quickly for off-diagonal elements as with our analysis, and that this will tend to hold for any implementation of the methodology using first, second, or third degree B-splines under some shrinkage.  Most test statistics will be practically independent and therefore $\alpha$ from inequality (\ref{simes}) nearly achieved. 

\section{Results}
\label{sec:results}
\subsection{Simulated data generation}

We generated two smooths with a specified degree of difference by first drawing $\{b_{k}\}$ with $k\in \{ 1,\dots , 120 \}$ from $N(0,\sigma_b^2 )$.  We then sampled 15, 20, or 30 indices, depending on the simulation, from $\{1, \dots ,120 \}$ and without replacement, which we called set $K$.  We sampled indices so that they tended to cluster around one another in order to be consistent with patterns of differences between two strata in reality.




For the $k$'s in set $K$, we drew non-zero $\Delta b_k$'s from $N(0,\sigma^2_\Delta)$, while for $k\notin K$, $\Delta b_k=0$.  We translated the non-zero $\Delta b_k$'s in the direction of their sign by value $M_\Delta$, assuring that all of the non-zero $\Delta b_k$'s were of size at least $M_\Delta$ in absolute value.  This allowed the $\Delta b_k$'s to exhibit some random variation, while encouraging uniformity in their magnitude to aid study.  


After generating the $\Delta b_k$'s, we took their sum with the corresponding $b_k$'s.  Call this complete set of 120 coefficients $\{ b_k^{(A)} \}$, the {\it altered} $b_k$'s, a different notation but analogous to ${\pmb b_{2}}$ above. 
Call the complete set of original $b_k$'s, unaltered by the $\Delta b_k$'s, $\{ b_k^{(U)} \}$, analogous to ${\pmb b_{1}}$ above.  So ${ b_k^{(A)}} - { b_k^{(U)}} = \Delta b_k =0$ for $k \notin K$, and ${ b_k^{(A)}} - { b_k^{(U)}}=\Delta b_k \neq 0$  for $k \in K$.  
The union of the support of the basis functions for which $\Delta b_k \neq 0$ is what we refer to below as the ``truly different region."

\subsection{Inference procedure}

We performed three kinds of simulations for continuous and binary outcomes, the first emulating real-world curves that a practitioner might want to model and used to create Figures \ref{eff_94}, \ref{eff_93}, \ref{eff_346}, and \ref{eff_338}.  The second kind of simulation varied the degree of difference between the two smooths and was used to create Figures \ref{test_group}, \ref{test_group2nd}, \ref{asymp_tdp_binary}, and \ref{tdp_tru_diff}.  Lastly, we performed simulations examining variability of actual TDP with respect to the TDP estimate for a fixed degree of difference $M_\Delta$, shown in Figures \ref{tdp_dist} and S2 and Table \ref{alpha_tab2}. We give parameters used for these simulations in Table \ref{param_summary}.

\begin{table}[ht!]
\centering
\caption{Parameterizations for the different simulation scenarios.}
\begin{tabular}{p{3.8cm}p{2.2cm}p{1.8cm}p{1.5cm}p{1.3cm}p{1.6cm}}  
  \hline
 Parameters & Proportion non-zero $\Delta b$ & Outcome & Sample size & Iterations & Figures \\ 
  \hline
  $\sigma_b^2 = 0.1$, $\sigma_\Delta^2 = 0.6$,\newline $\alpha = 0.15$, $M_\Delta = 0.93$, $\sigma^2 = 0.8$  & 0.125 and 0.25 & Continuous & 4000 per stratum & 1 &  1, 2  \\ 
  $\sigma_b^2 = 0.1$, $\sigma_\Delta^2 = 0.6$,\newline $\alpha = 0.15$, $M_\Delta = 3.4$, $\sigma^2 = \mbox{N/A}$  & 0.167 & Binary & 4000 per stratum & 1 & 3, S1 \\ 
  $\sigma_b^2 = 0.1$, $\sigma_\Delta^2 = 0.05$,\newline $\alpha = 0.2$, $M_\Delta=(0,2.5)$, $\sigma^2 = 0.8$  & 0.125 and 0.25 & Continuous & 4000 per stratum & 1000 & 4, 6\\ 
  $\sigma_b^2 = 0.1$, $\sigma_\Delta^2 = 0.05$,\newline $\alpha = 0.2$, $M_\Delta=(0,9)$, $\sigma^2 =  \mbox{N/A}$  & 0.167 & Binary & 4000 per stratum & 1000 & 7\\ 
 $\sigma_b^2 = 0$, $\sigma_\Delta^2 = 0.005$,\newline $\alpha = \{0.1,0.2,0.3\}$, $M_\Delta = 2.15$, $\sigma^2 = 0.8$  & 0.125 and 0.3 & Continuous & 4000 per stratum & 1000 & 5, S2  \\ 
 $\sigma_b^2 = 0.1$, $\sigma_\Delta^2 = 0.6$,\newline $\alpha = 0.2$, $M_\Delta=(0,2.5)$, $\sigma^2 = 0.8$  & 0.125 and 0.3 & Continuous & 4000 per stratum & 1000 & S3 \\ 
   \hline
\end{tabular}
\label{param_summary}
\end{table}







To explain some of the parameterization decisions, we decreased $\sigma_\Delta^2 = 0.05$ in a subset of simulations so that the magnitude of non-zero $\Delta b_k$'s would be dominated by $M_\Delta$, ensuring greater uniformity in them.  This helped isolate effects of the $M_\Delta$ parameter on TDP estimation.  For the variable $M_\Delta$ simulations, we increased $M_\Delta$ from 0 to 2.5 linearly over the 1000 iterations for the continuous outcome case.   Simulations for binary data are underpowered as compared to the continuous case and so we varied $M_\Delta$ between 0 and 9 over the 1000 iterations for that scenario.  In each case, for every iteration a different collection of $\{b_k \}$ and set $K$ were generated.  The model noise, $\sigma^2$, is not applicable in the binary outcome case because one draws from the Bernoulli distribution with success probability determined by the inverse logit of the linear predictor. That linear predictor was distributed Gaussian, being a sum of Gaussian-distributed basis function parameters.  When in figures and tables we use labels of 0.5, 0.7, or 0.9 TDP, these were arrived at by ordering p-values and estimating TDP progressively on regions corresponding to that ordering until one reached the specified level.  Smooths were fit to each stratum using the {\it mgcv} package in R and the tuning parameter chosen with restricted maximum likelihood, while we calculated equations \ref{phi_eqn} and \ref{h_eqn} of the algorithm with package {\it hommel} \citep{wood_fast_2011,goeman_hommel_2021}.

\subsection{Simulation results}
\label{simulation}

\begin{figure}[H]
\begin{center}
\includegraphics[width=\textwidth]{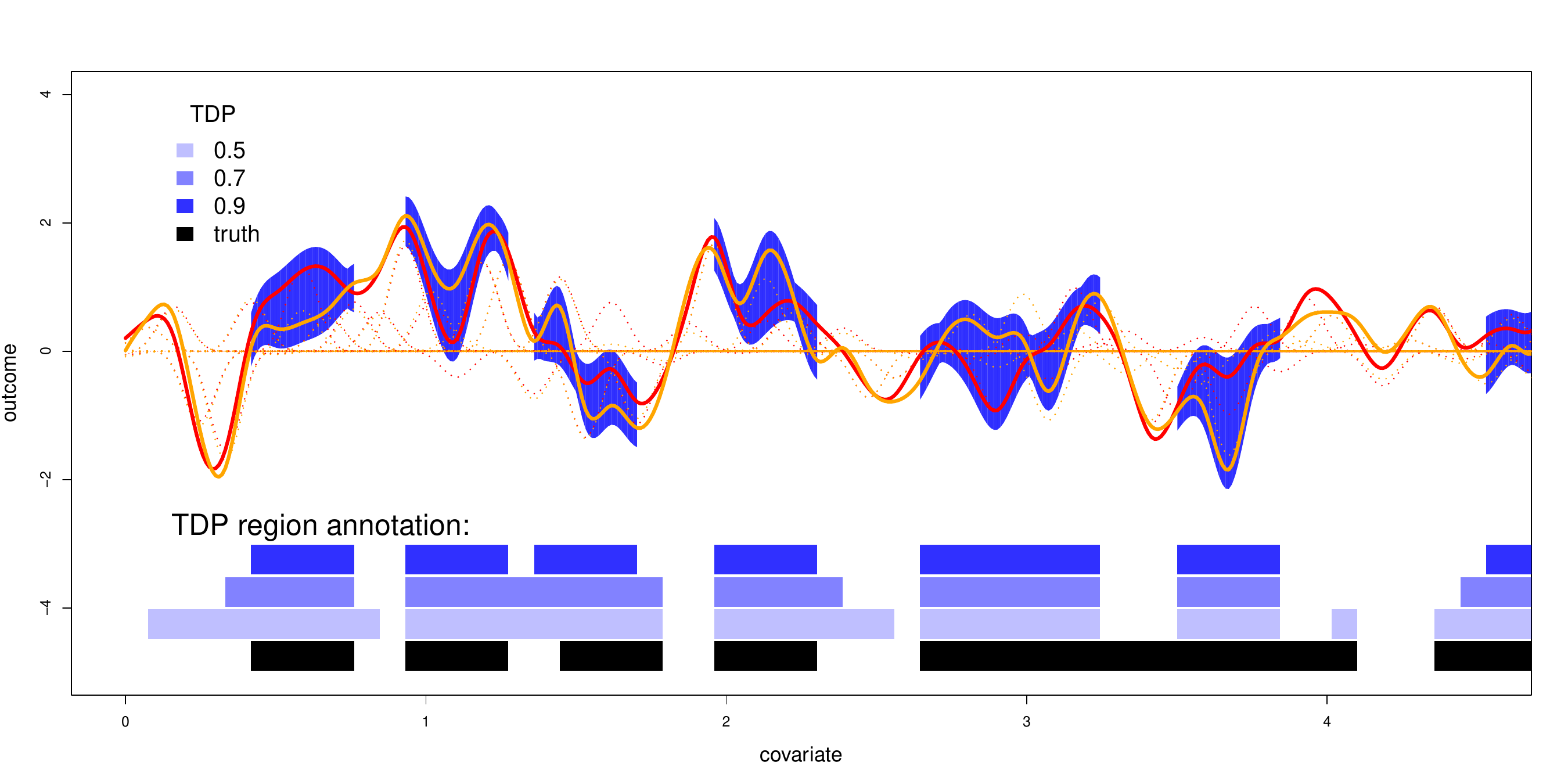} 
\caption{Two simulated curves on a domain of (0,4.5) with highlighted difference regions along the smooths in dark blue.  The highlighted dark blue region corresponds to the 0.9 TDP annotation of the same color shown at the bottom of the figure.  There is analogous `bar' annotation for estimated TDP's of 0.7 and 0.5 in different shades of blue.  The black region at the most bottom shows the intervals where the 2 curves are generated from different basis functions.  The many dotted line curves in red and orange show the underlying basis functions scaled according to the true basis coefficients.  The estimates of their superimpositions -- the estimated smooth -- are the thicker, solid red and orange curves.  We see that the TDP region annotation of 0.5, 0.7, and 0.9, are accurate estimates of TDP as compared to the truth, also confirmed in Figure \ref{tdp_dist}.  This figure corresponds to a minimum effect size delta of 0.94 in the difference regions.}
\label{eff_94}  
\end{center}
\end{figure}

\begin{figure}[H]
\begin{center}
\includegraphics[width=\textwidth]{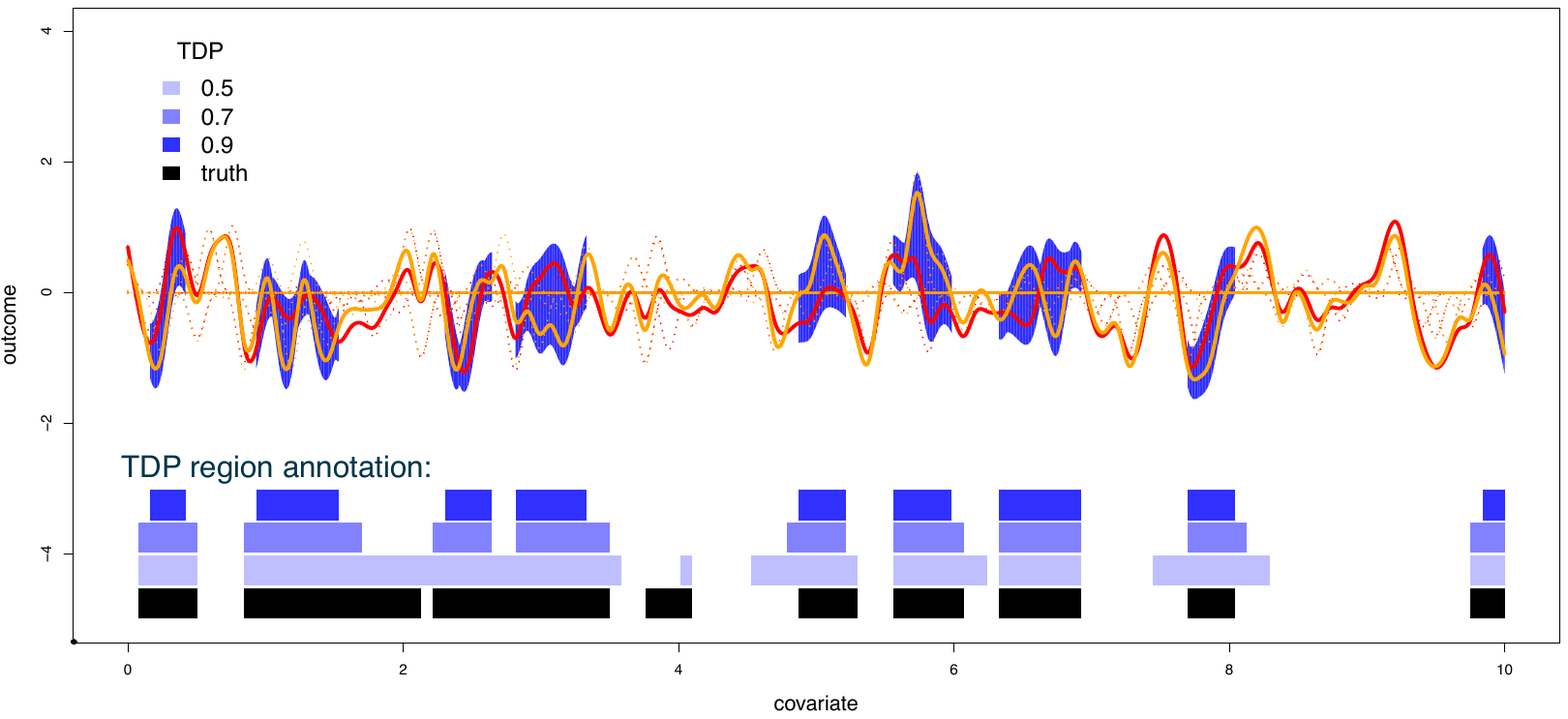} 
\caption{Two simulated curves on a domain of (0,10) with highlighted difference regions along the smooths in dark blue.  The highlighted dark blue region corresponds to the 0.9 TDP annotation of the same color shown at the bottom of the figure.  There is analogous `bar' annotation for estimated TDP's of 0.7 and 0.5 in different shades of blue.  The black region at the most bottom shows the intervals where the 2 curves are generated from different basis functions.  The many dotted line curves in red and orange show the underlying basis functions scaled according to the true basis coefficients.  The estimates of their superimpositions -- the estimated smooth -- are the thicker, solid red and orange curves.  We see that the TDP region annotation of 0.5, 0.7, and 0.9, are accurate estimates of TDP as compared to the truth, also confirmed in Figure \ref{tdp_dist}.  This figure corresponds to a minimum effect size delta of 0.93 in the difference regions.}
\label{eff_93} 
\end{center}
\end{figure}

\begin{figure}[H]
\begin{center}
\includegraphics[width=\textwidth]{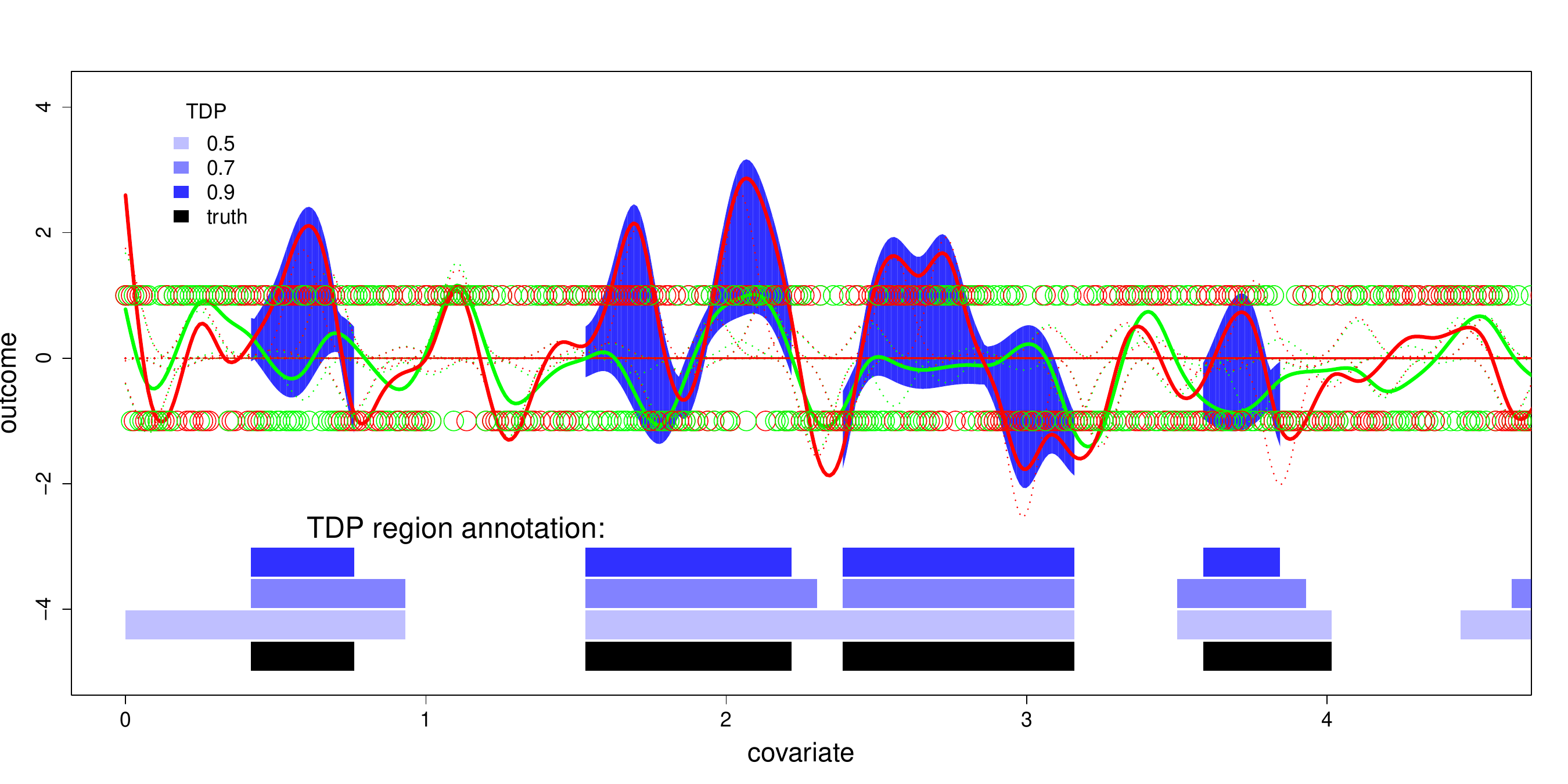} 
\caption{Two simulated curves for a binary outcome on a domain of (0,4.5) with highlighted difference regions along the smooths in dark blue.  The binary outcome was drawn from an inverse-logit transformed linear predictor, that distributed Gaussian being a linear combination of Gaussian random variables.  The data are fit using logistic regression and the smooths shown correspond to the estimated linear predictors.  The points plotted along 1 and -1 of the y-axis are the modelled 1's and 0's, respectively, colored according to the corresponding smooth, and drawn on the graph in a random order so that their shade of color communicates the relative quantity of each.  The highlighted dark blue region corresponds to the 0.9 TDP annotation of the same color shown at the bottom of the figure.  There is analogous `bar' annotation for estimated TDP's of 0.7 and 0.5 in different shades of blue.  The black region at the most bottom shows the intervals where the 2 curves are generated from different basis functions.  The many dotted line curves in red and orange show the underlying basis functions scaled according to the true basis coefficients.  The estimates of their superimpositions -- the estimated linear predictor smooths -- are the thicker, solid red and orange curves.  We see that the TDP region annotation of 0.5, 0.7, and 0.9, are relatively accurate estimates of TDP as compared to the truth, also confirmed in Figure \ref{asymp_tdp_binary}.  This figure corresponds to a minimum effect size delta of 3.46 in the difference regions.}
\label{eff_346}  
\end{center}
\end{figure}

\begin{figure}[H]
\centering
\begin{subfigure}{0.4\textwidth}
\includegraphics[width=\linewidth]{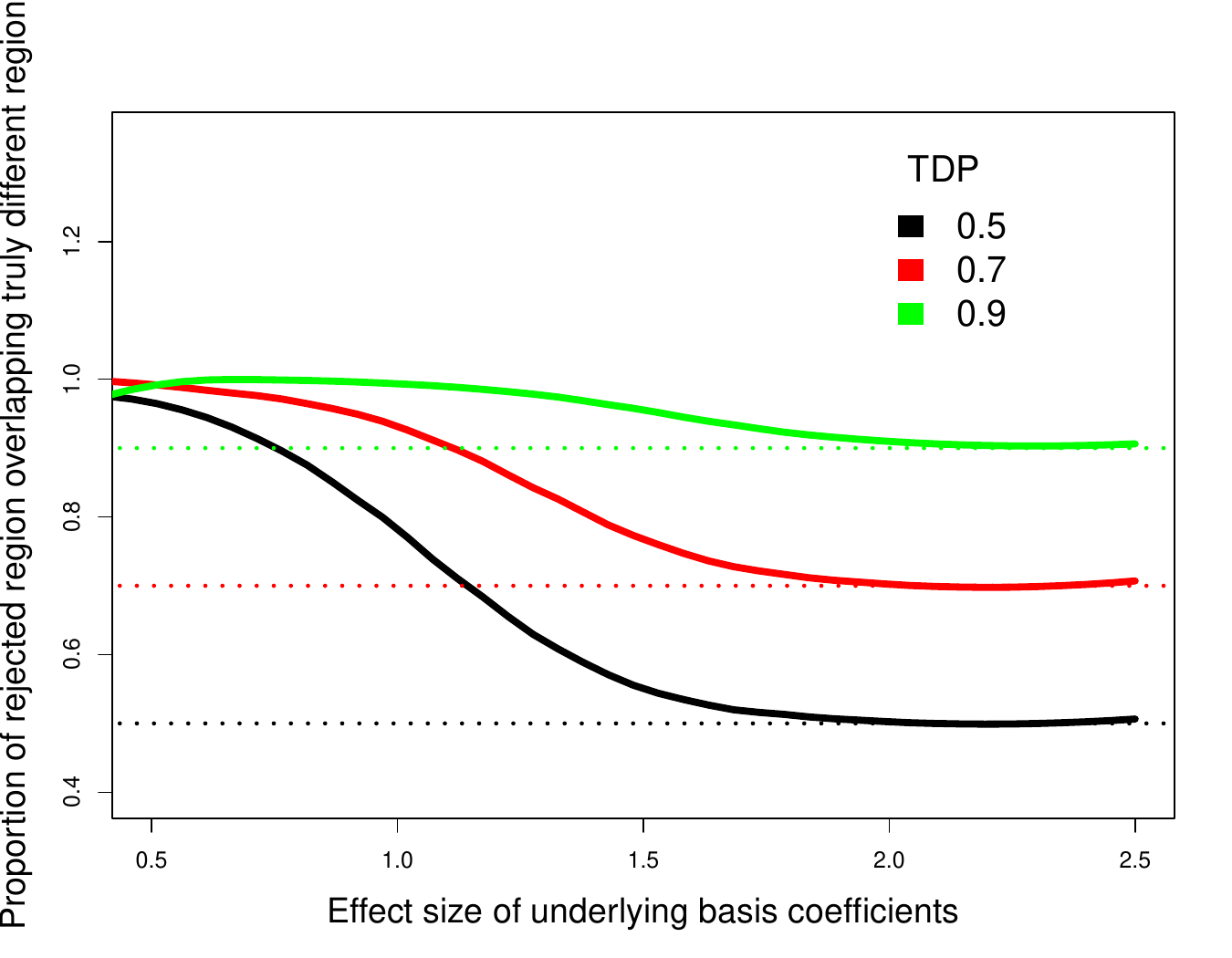}
\caption{15 of 120 non-zero $\Delta b_k$'s} 
\label{test_group1}
\end{subfigure}
\hspace*{1.5cm} 
\begin{subfigure}{0.4\textwidth}
\includegraphics[width=\linewidth]{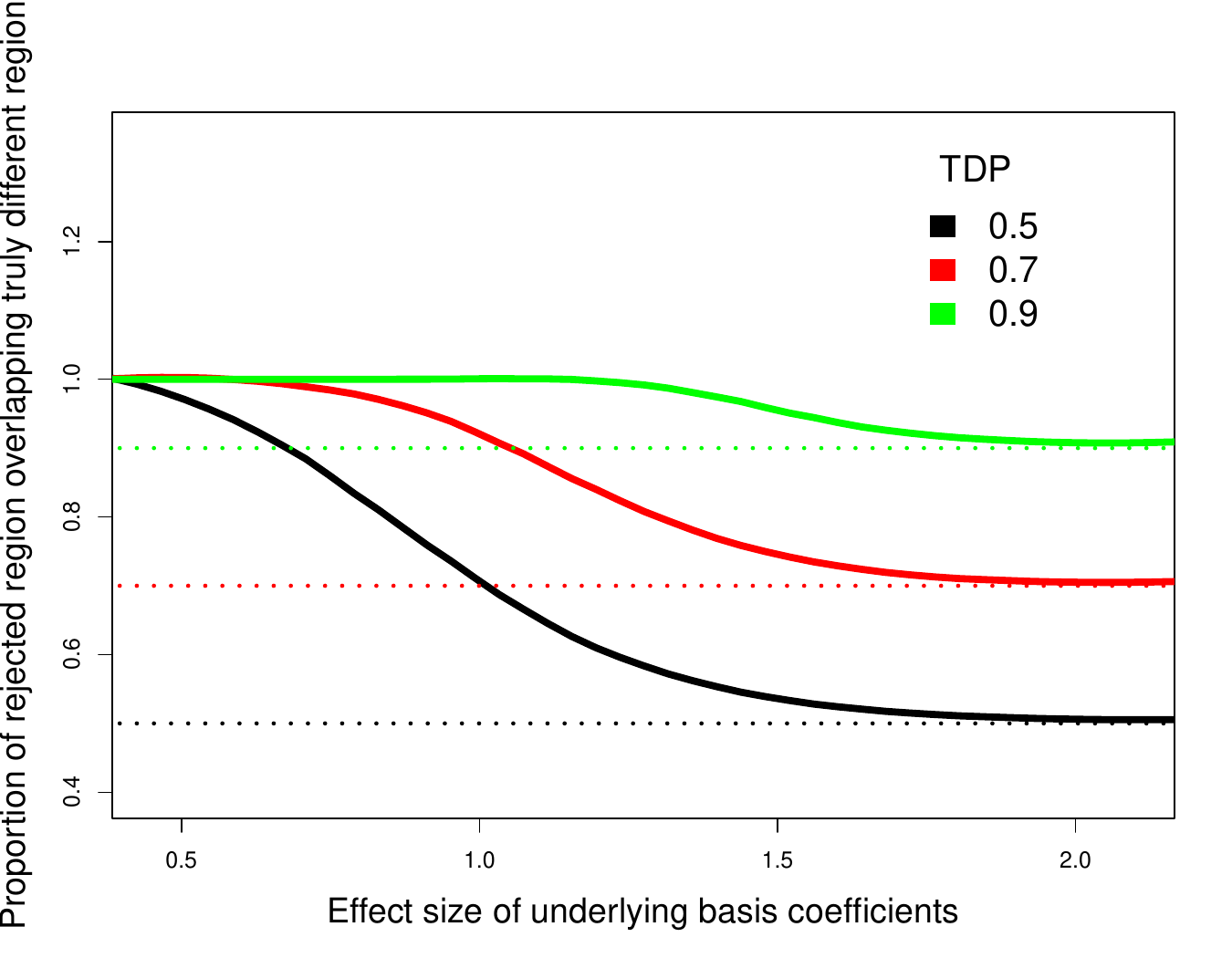}
\caption{30 of 120 non-zero $\Delta b_k$'s} 
\label{test_group2}
\end{subfigure}
\captionsetup{width=0.8\textwidth}
\caption{Actual TDP for regions when using estimates of 0.5, 0.7, and 0.9, shown in green, red, and black, respectively.  Lines are loess smooths calculated from approximately 1000 simulations over the different TDP thresholds.  Each single underlying data point was calculated as a function of the overlap of the TDP annotation bars versus truth, see also Figures \ref{eff_94}, \ref{eff_93}, and \ref{eff_346} as examples. } 
\label{test_group}
\end{figure}

\begin{figure}[H]
\centering
\begin{subfigure}{0.31\textwidth}
\includegraphics[width=\linewidth]{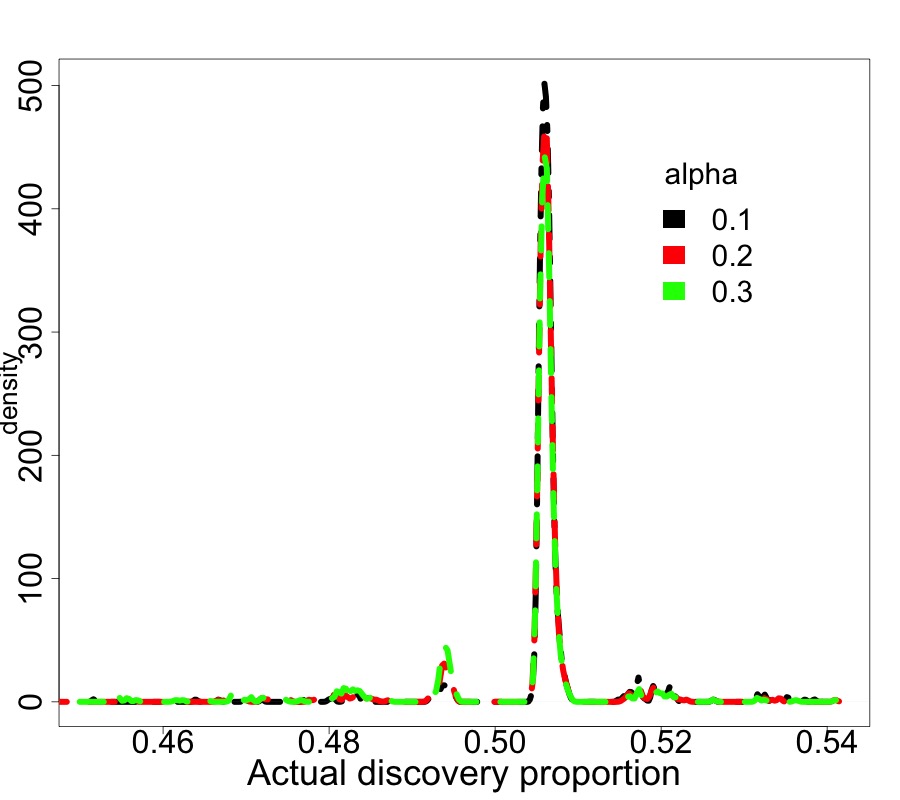}
\caption{Actual TDP for an estimate of 0.5} 
\label{tdp_dist1}
\end{subfigure}
\begin{subfigure}{0.31\textwidth}
\includegraphics[width=\linewidth]{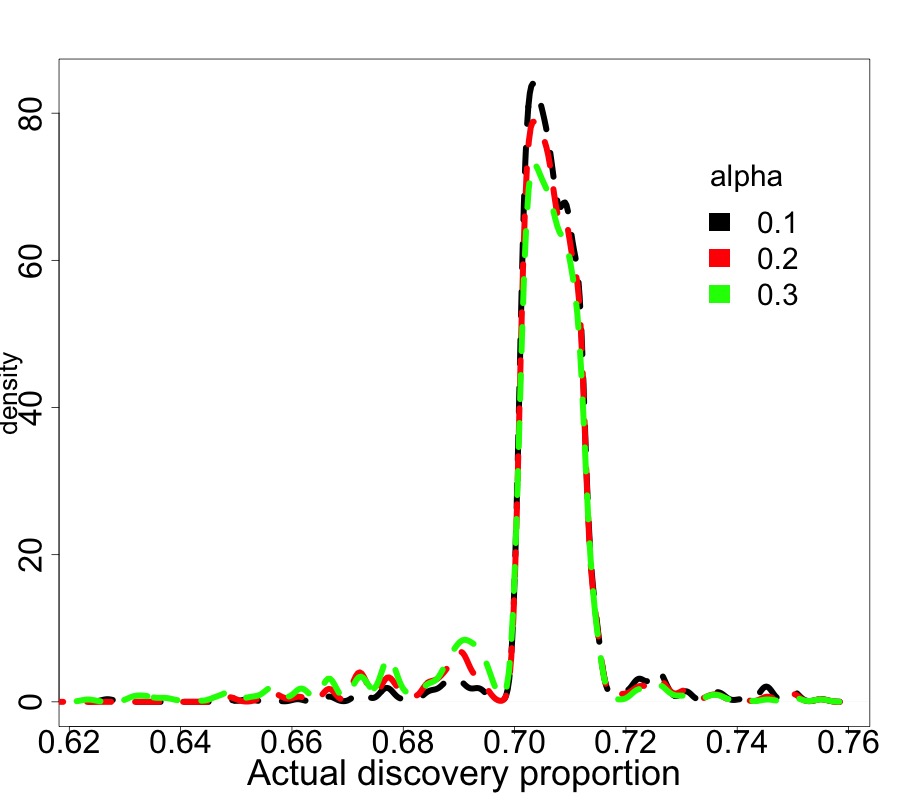}
\caption{Actual TDP for an estimate of 0.7} 
\label{tdp_dist2}
\end{subfigure}
\begin{subfigure}{0.31\textwidth}
\includegraphics[width=\linewidth]{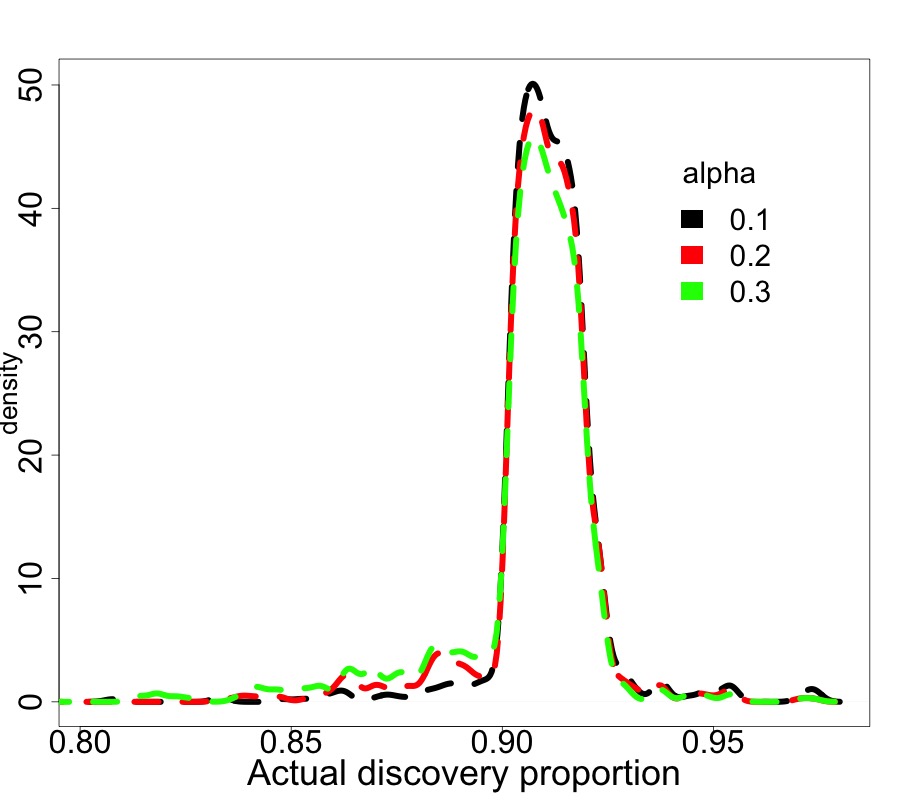}
\caption{Actual TDP for an estimate of 0.9} 
\label{tdp_dist3}
\end{subfigure}
\captionsetup{width=0.8\textwidth}
\caption{Distribution of actual TDP over 1000 iterations with 15 non-zero $\Delta b_k$'s for estimated TDPs of a) 0.5, b) 0.7, and c) 0.9, under different $\alpha$'s, specified in the legend.  Distributions were estimated using a kernel density estimator.  One observes all distributions to be relatively narrow.  One sees slightly lower density around each distribution's mode for increasing $\alpha$, and slightly higher density around the tails.}
\label{tdp_dist}
\end{figure}

\begin{figure}[H]
\centering
\begin{subfigure}{0.31\textwidth}
\includegraphics[width=\linewidth]{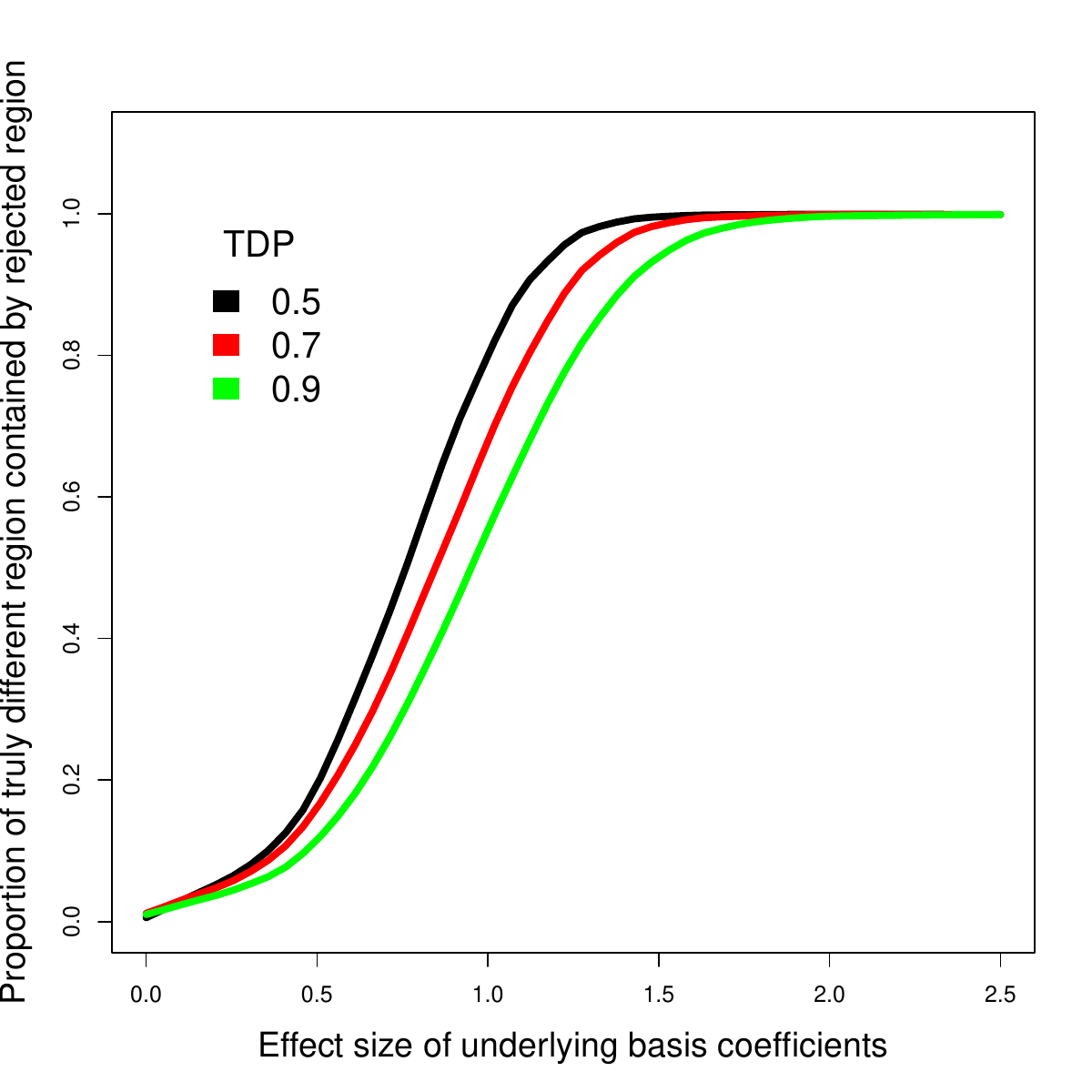}
\caption{15 of 120 non-zero $\Delta b_k$'s} 
\end{subfigure}
\hspace*{1.5cm} 
\begin{subfigure}{0.31\textwidth}
\includegraphics[width=\linewidth]{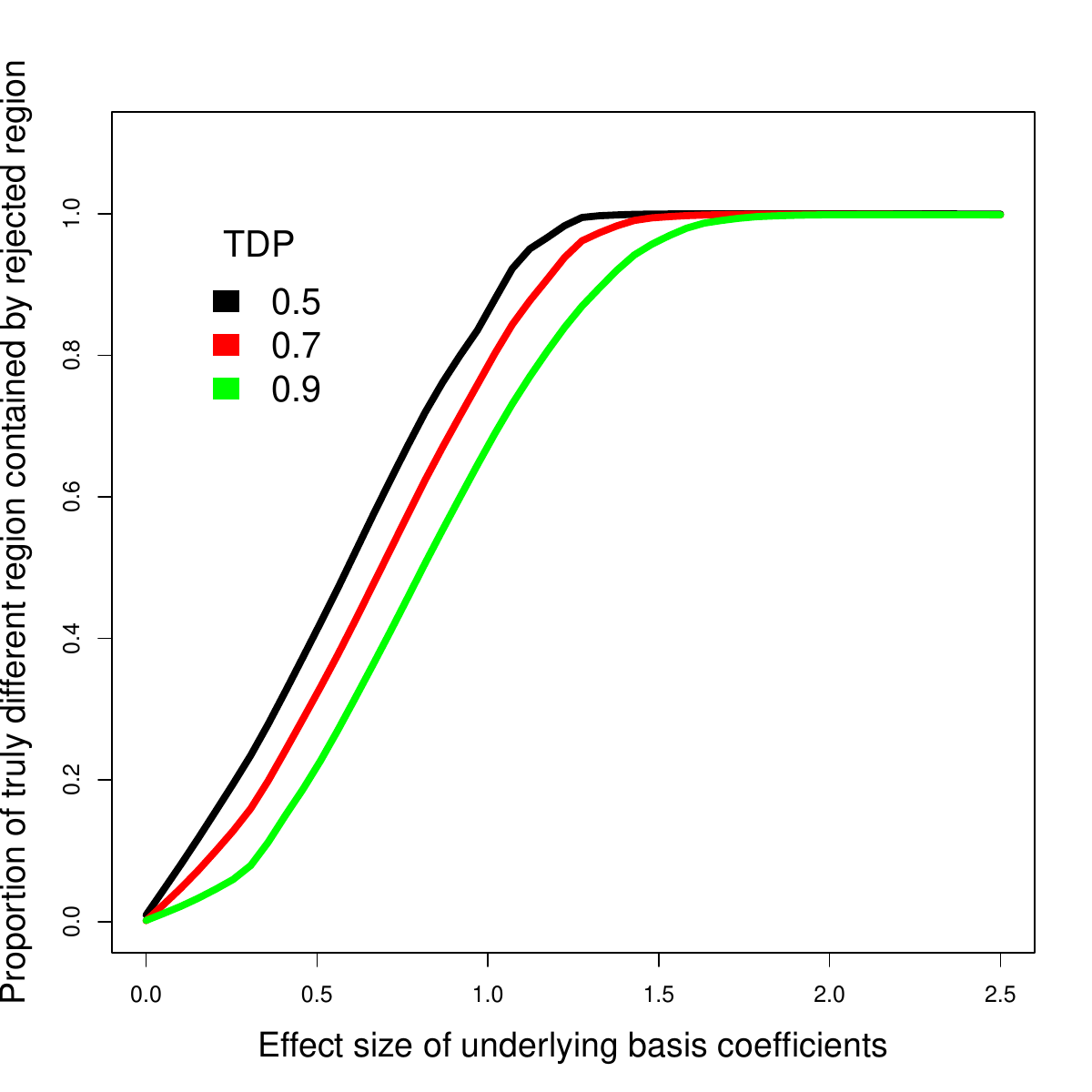}
\caption{30 of 120 non-zero $\Delta b_k$'s} 
\end{subfigure}
\captionsetup{width=0.8\textwidth}
\caption{Proportion of the truly different region labelled as such at different TDP estimate thresholds for increasing $\Delta b_k$. }
\label{test_group2nd}
\end{figure}

\begin{figure}[H]
\begin{center}
\includegraphics[width=0.5\textwidth]{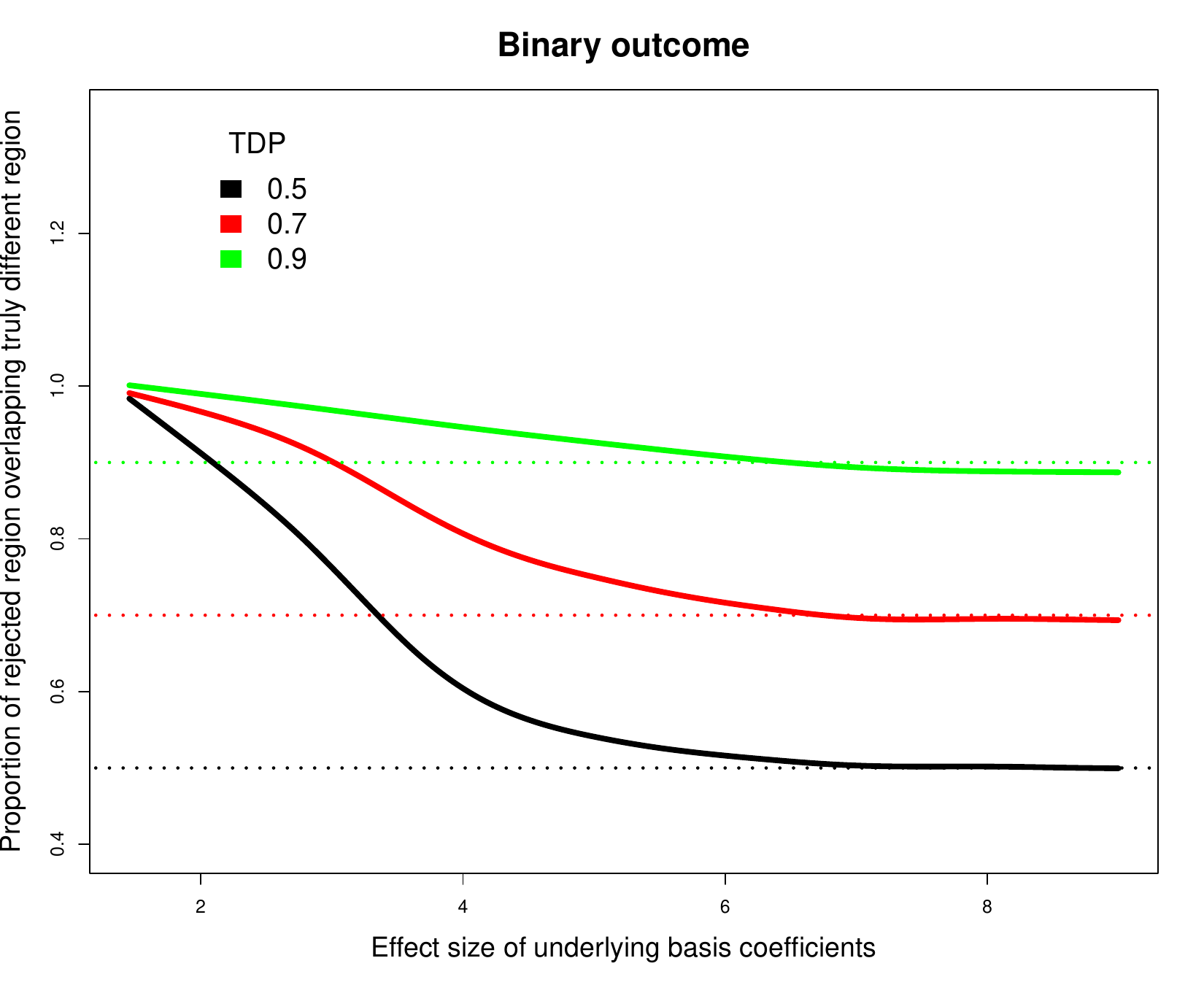}
\captionsetup{width=0.7\textwidth}
\caption{Average TDP over a range of minimum effect size differences $M_\Delta$ for $20$ non-zero $\Delta b_k$'s when modelling the binary outcome, at thresholds of 0.5, 0.7, and 0.9 and using an $\alpha=0.2$.}
\label{asymp_tdp_binary}
\end{center}
\end{figure}

We examined different aspects of TDP estimation in each simulation, shown in Figures \ref{eff_94}--\ref{asymp_tdp_binary}, S1-S3 and Table \ref{alpha_tab2}. 
These calculations can be categorized into four types: 

\begin{enumerate}

\item[{\bf (1)}] Simulated smooths and their true regions of difference alongside annotation of different TDP estimates using our methodology, Figures \ref{eff_94}--\ref{eff_346}, S1

\item[{\bf (2)}] Actual TDP's approaching their estimated level (intersection of highlighted TDP regions of Figures \ref{eff_94}--\ref{eff_346} and the truly different region, divided by size of the TDP regions), Figures \ref{test_group} and \ref{asymp_tdp_binary} (varying $\Delta b$), Figures \ref{tdp_dist} and S2 and Table 2 (fixed $\Delta b$).  This proportion should align with the estimated TDP.

\item[{\bf (3)}] Proportion of the truly different region covered by the TDP region (intersection of highlighted TDP regions of Figures \ref{eff_94}--\ref{eff_346} and truth, divided by size of truth), Figure \ref{test_group2nd}

\item[{\bf (4)}] Estimated TDP of the region simulated as truly different, Figure \ref{tdp_tru_diff}

\end{enumerate}



Figures \ref{eff_94}--\ref{eff_346} reveal our methodology performs as intended.  The two smooths in each figure depicted with thick lines are those estimated from the simulated data, while the dotted lines of the respective colors in each figure are basis functions scaled by the underlying simulated coefficients. The highlighted region along the smooths is the 0.9 TDP region and corresponds to that depicted with rectangles at the bottom of the figure.  The bottom of the figure also shows the 0.5 and 0.7 TDP regions in different shades of blue, and those regions of the two smooths which were simulated as truly different  (i.e., $\Delta b_k \neq 0$ implies $b_k^{(A)}\neq b_k^{(U)}$) in black.  Since the simulated smooths are those estimated from the data, visual inspection of their differences may not align identically with the truly different region shown in black.
Visual inspection of TDP regions of 0.5, 0.7, and 0.9, show them overlapping the truly different region in approximately the proportion of the estimate, explored in a more precise way in Figures \ref{test_group}, \ref{tdp_dist}, \ref{asymp_tdp_binary}, and Table \ref{alpha_tab2}. 


$$\mbox{[  Figures \ref{eff_94} and \ref{eff_93}  ]}$$

Interpretation of Figure \ref{eff_346} is similar to Figures \ref{eff_94} and \ref{eff_93}, except based on binary data so that the estimated smooths depict the linear predictor.  The $1$ and $0$ outcomes are mapped to $1$ and $-1$, respectively, in the figure, and in colors correspondent with the smooths to give a sense of the relative quantity of these data points in the two modelled strata and their influence on the estimated smooth linear predictor.  Figure \ref{eff_338} in the Supplementary Material can be understood in the same way.  

$$\mbox{[  Figure \ref{eff_346}  ]}$$

Figure \ref{test_group} shows what proportion of the 0.5, 0.7, and 0.9 TDP regions cover the truly different region under increasing effect size differences ($\Delta b_k$). 
The loess fit lines in the figures were based on the 1000 simulations.  One sees that for all effect sizes and proportions of non-zero $\Delta b_k$'s, the actual TDP falls above or on its estimated value (estimated TDP value shown as dotted lines of the corresponding green, red, and black colors), with convergence of actual TDP to estimated TDP for increasing effect size.  


$$\mbox{[  Figure \ref{test_group}  ]}$$

The distribution of different estimated TDPs for different $\alpha$'s shown in Figure \ref{tdp_dist} reveals the tight distribution of the estimate around actual TDP in all cases.  Table \ref{alpha_tab2}a likewise shows estimates performing as intended at all $\alpha$'s and estimated TDPs of 0.5, 0.7, and 0.9. Figure \ref{test_group2nd} shows the proportion of the truly different region labelled as such at different TDP estimates for increasing $\Delta b_k$, where 
one observes curves for all TDP levels tending to be lower or right-shifted for smaller numbers of non-zero $\Delta b_k$'s.

$$\mbox{[  Figures  \ref{tdp_dist} and \ref{test_group2nd}  ]}$$

Figure \ref{asymp_tdp_binary}, generated from the binary outcome, shows actual TDP starting above and then approaching the estimated, or nominal, TDP for increasing effect size differences, though at higher values than observed for the continuous outcome.  These estimates tend to be very close to their actual values at all levels of estimated TDP.


$$\mbox{[  Figure  \ref{asymp_tdp_binary}  ]}$$




\section{Data analysis}
\label{data_analy}

\begin{figure}[H]
\begin{center}
\includegraphics[width=0.8\textwidth]{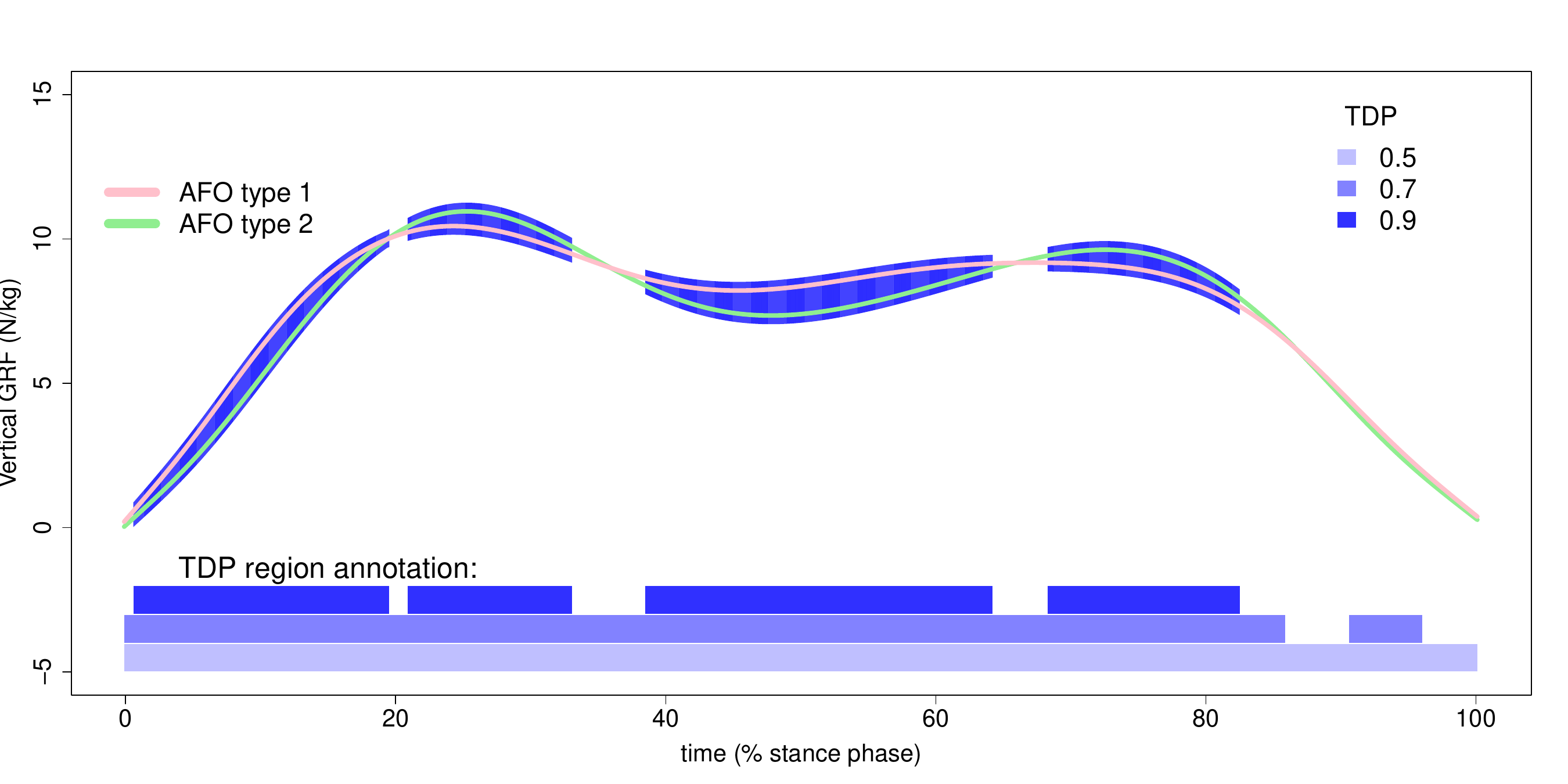}
\captionsetup{width=0.85\textwidth}
\caption{
AFO type 1 and type 2 ground reaction force smooths, plotted against \% stance phase for a single stride.  Highlighted annotation on the curves is the 90\% TDP difference region with simultaneous confidence using $\alpha= 0.01$, and aligns with the block annotation of the same color shown at the bottom of the figure. Annotation at the bottom of the figure also shows 70\% and 50\% TDP regions of difference in the two smooths, colored according to that shown in the legend.
}
\label{gait}
\end{center}
\end{figure}

We analyzed a study of walking gait in 24 post-surgical intervention children with bilateral spastic cerebral palsy who received one of two ankle-foot orthosis (AFO) types to improve stability during stride, a subset of a study cohort described in \citet{skaaret_impact_2019}. These two interventions being compared consisted of approximately 3,400 and 5,200 vertical ground reaction force (GRF, measured in Newtons/kg)  by  percent  stance  phase  data  points measured on 10 and 14 children, respectively,  where  100\%  corresponds  to  a completed stride.  Each patient had approximately 350 GRF observations for a single stride, with the fewest being 270 and the most 526.  The large number of GRF measurements per patient gave opportunity to model smooths with a large number of knots and assess on a fine grain where patients’ strides differed based on the type of orthotic used.  With the analysis we hoped to answer how similar was GRF in the valleys (the region of stride between the two GRF peaks) of the two AFO groups.  We used 150 identically placed knots for the AFO type 1 and type 2 population average smooths.  Within-person correlation was controlled for using subject-specific smooths, each on 3 knots to avoid the excessive flexibility that would make inference on the orthotic intervention difficult.  
The analysis revealed that the 0.9 TDP region covered most of the valley region of patients' strides, with the exception of a small area on either side of it where the two subgroups' GRF smooths crossed (Figure \ref{gait}).  The interpretation of the result is that the proportion of true difference in the 0.9 TDP region is at least 0.9 with simultaneous $\alpha= 0.01$.  The 0.7 TDP region covered nearly the entirety of the stride with the exception of the approximate last 15\%.

$$\mbox{[  Figure \ref{gait}  ]}$$

\section{Discussion and Extensions}
\label{sec:discussion}

We described a procedure to make true discovery proportion statements on regions where two smooths differ.  We used a closed testing procedure on parameter estimates underlying smooths and leveraged recent results on TDP confidence statements for Simes' local tests so that the method has low computational cost.  The procedure gives simultaneous strong control of the family wise error rate, and 
the simultaneous error bounds allow an analyst to posthoc test additional hypotheses based on preliminary exploration and avoid inflating error rates. And it relies on the multivariate $\chi^2$ test statistics being positive regression dependent on subsets, which we showed, in addition to the compact support of 
B-spline basis functions. 



There are several extensions of this work that render it applicable to other settings.  The methodology is valid for any generalized linear model, and we provided simulations for binary outcomes demonstrating its applicability in that setting (Figures 3, 7, S1).  
One can also extend the procedure to testing differences in more than two smooths.  The extension is straightforward since smooths are estimated independently so one need only modify the hypothesis tests comparing groups of relevant basis coefficients.  
One could apply multivariate analysis of variance methods such as Wilk's lambda to the coefficient vectors underlying smooths or design Wald-type test statistics using parameter contrasts and subsequently perform TDP estimation with calculated p-values \citep{mardia_multivariate_1979}. We explore the latter of these options in Figure \ref{fig:3strat} which demonstrates accurate TDP estimate performance.Application to small samples on only two smooths is also feasible with closed-form formulae for estimation of the model penalization parameter and use of Hotelling's T$^2$ \citep{wand_miscellanea_1999,hotelling_generalization_1931}. 

\section{Acknowledgments}

The author thanks Dr.$\,$Ingrid Skaaret and the Department of Child Neurology at Oslo University Hospital for use of the data and Prof.$\,$Arnoldo Frigessi for pointing to helpful multiple testing literature.  The research was partially supported by the National Institutes of Health grant 1P01CA261669.


\section{Supporting Material}

Additional information for this article is available online. It contains proofs of Theorems  2 and 4, Lemmas 1 and 3, and additional simulation results, found in Appendices S1-S4.

\bibliographystyle{plainnat} 
\bibliography{refs_fdr_splines1_6_23}


\clearpage

\newgeometry{left=0.75cm,right=1.4cm}

\section{Figures and Tables}

\begin{table}[!htb]
    \caption{Actual TDP's for varying $\alpha$ levels and TDP estimates.  Different subtables correspond to a varying number of non-zero $\Delta b_k$'s.}
    \begin{minipage}{.48\linewidth}
\captionsetup{width=0.8\textwidth}
      \centering
       \caption*{a) 15 of 120 non-zero $\Delta b_k$'s. Proportions calculated on 1000 simulations over the different TDP thresholds.}
  \small
\begin{tabular}{rrrr}
  \hline
 & TDP=0.5 & TDP=0.7 & TDP=0.9 \\ 
  \hline
alpha=0.1 & 0.506 & 0.707 & 0.911 \\ 
alpha=0.2 & 0.505 & 0.705 & 0.909 \\ 
alpha=0.3 & 0.504 & 0.703 & 0.906 \\ 
   \hline
\end{tabular}
    \end{minipage}%
    \begin{minipage}{.48\linewidth}
    \captionsetup{width=0.8\textwidth}
      \centering
        \caption*{b) 30 of 120 non-zero $\Delta b_k$'s. Proportions calculated on 1000 simulations over the different TDP thresholds.}
  \small
\begin{tabular}{rrrr}
  \hline
 & TDP=0.5 & TDP=0.7 & TDP=0.9 \\ 
  \hline
alpha=0.1 & 0.507 & 0.706 & 0.909 \\ 
alpha=0.2 & 0.505 & 0.703 & 0.906 \\ 
alpha=0.3 & 0.503 & 0.701 & 0.903 \\ 
   \hline
\end{tabular}


    \end{minipage} 
    \label{alpha_tab2}
\end{table}







\restoregeometry

\clearpage

\begin{center}
\textbf{\large Supplementary Material: TDP perspective on localizing differences in smooths}

\textbf{David Swanson}
\end{center}
\setcounter{equation}{1}
\setcounter{figure}{0}
\setcounter{section}{0}
\setcounter{table}{0}
\setcounter{page}{1}
\makeatletter
\renewcommand{\thetable}{S\arabic{table}}
\renewcommand{\thefigure}{S\arabic{figure}}
\renewcommand{\thesection}{S\arabic{section}}
\renewcommand{\bibnumfmt}[1]{[S#1]}
\renewcommand{\citenumfont}[1]{S#1}
\newcommand{\ma}[1]{\mathbf{#1}}

\section{PRDS on multivariate $\chi^2$ of generalized Wishart type}

{\bf Proof of Theorem 2}

\noindent
{\bf Note:} {\it We adopt a different notation to that in Definition 1, with $U_{ik}$ in that definition analogous to $x^*_{ik}$ in this proof and ${\pmb x_{i}}$ analogous to $({\hat{\pmb b}}_1- {\hat{\pmb {b}}}_2)_{i,d}$. $Q_i$ is used in the same way in both the definition and here.  
Also, $v_i$ in Definition 1, the dimension of multivariate Gaussian random variable $(U_{i,1} \dots U_{i,v_i})$, is $d_i$ here, whereas $v_i$ here takes the role of eigenvectors. }


Consider ${\pmb x_i}= (x_{i,1} \dots x_{i,d_i})$ for $i=1\dots m$ with ${\pmb x_i}\sim MVN(0, \Sigma_{ii})$, the multivariate Gaussian distribution with mean 0 and covariance $\Sigma_{ii}$, and suppose between any two ${\pmb x_i}$ and ${\pmb x_j}$ we have $Cov({\pmb x_i} , {\pmb x_j}) = \Sigma_{ij}$. 
Because for the TDP estimation procedure we ultimately use test statistics ${\pmb x_{i}}' \, \Sigma_{ii}^{-1} \, {\pmb x_{i}}$ for all $i$, on which PRDS must be shown for the collection, we will focus on transformed ${\pmb x_i}^*=({\pmb x_{i,1}^*} \dots {\pmb x_{i,d_i}^*})'$, $i=1\dots m$, which is $MVN(0, I)$, $I$ the identity matrix, where $Cov(x_i^*,x^*_j)=\Sigma_{ii}^{-1/2} \Sigma_{ij} \Sigma_{jj}^{-1/2}$. Call this covariance $A_{ij}=\Sigma_{ii}^{-1/2} \Sigma_{ij} \Sigma_{jj}^{-1/2}$, and note that $A_{ij}$ may have elements of 1, relevant because of the `sliding window' used in our case in constructing test statistics prior to TDP estimation. Now define $Q_i = \sum_{k=1}^{d_i} x_{i,k}^{*2}$ for all $i$, and notice $Q_i \sim  \chi^2_{d_i}$, the central $\chi^2$ distribution on $d_i$ degrees of freedom.  To show $(Q_1, \dots , Q_m)$ is PRDS, it is enough that every component in the conditional distribution $(Q_1 , \dots , Q_{j-1} , Q_{j+1} , \dots , Q_m) \, |\, Q_j $ 
be stochastically increasing as a function of increasing $Q_j$. 




To show this, first consider the conditional distribution $({\pmb x_{1}^*}, \dots ,{\pmb x_{j-1}^*},{\pmb x_{j+1}^*} \dots {\pmb x_{m}^*} ) \,| \, {\pmb x_{j}^*}$.  Using properties of the conditional multivariate Gaussian, $({\pmb x_{1}^*}, \dots ,{\pmb x_{j-1}^*},{\pmb x_{j+1}^*} \dots {\pmb x_{m}^*} ) \,| \, {\pmb x_{j}^*}$ are jointly multivariate Gaussian and any component ${\pmb x_{i}^*}$ in this conditional distribution is distributed $MVN( \mu_{i,x_j^*}\,,\, I - A_{ij} A_{ij}')$, where $\mu_{i,x_j^*} = A_{ij} \, {\pmb x_{j}^*}$. 




Now consider $f({\pmb x^*_j} |\, Q_j = l)$, the conditional density of ${\pmb x^*_j}$ given $Q_j = l$, and recall ${\pmb x^*_j} = (x^*_{j,1} \dots x^*_{j,d_j} )$ and $Q_j =  \sum_{k=1}^{d_j} x_{j,k}^{*2}$.  Marginally, ${\pmb x^*_j}\sim MVN(0, I)$.  Because the covariance is identity, this conditional density $f({\pmb x^*_j} |\, Q_j = l)$ is some constant $k$ for all $x^*_{j,1} , \cdots ,  x^*_{j,d_j}$ satisfying $x^{*2}_{j,1} + \cdots + x^{*2}_{j,d_j}=Q_j=l$ because the ${\pmb x^*_{j}}$ satisfying $x^{*2}_{j,1} + \cdots + x^{*2}_{j,d_j}=l$ defines a hypersphere of radius $\sqrt{l}$ in ${\mathbb R}^{d_j}$ centered at 0, and that hypersphere follows a constant-density contour of the multivariate Gaussian when the Gaussian is centered at $0$ and has identity covariance.   
Call the set of entries of ${\pmb x_j^*}$ consistent with the conditioning event $B_l$, the surface of the $d_j$-sphere of radius $\sqrt{l}$ centered at 0. 



Consider again the conditional distribution $({\pmb x_{1}^*}, \dots ,{\pmb x_{j-1}^*},{\pmb x_{j+1}^*} \dots {\pmb x_{m}^*} ) \,| \, {\pmb x_{j}^*}$, and let us focus on one of its components, say ${\pmb x_{i}^*} $, so that we're considering the distribution of ${\pmb x_{i}^*} \,| \, {\pmb x_{j}^*}$, which again is multivariate Gaussian. 
As written, the covariance of ${\pmb x_{i}^*} \,| \, {\pmb x_{j}^*}$ is $I-A_{ij}A_{ij}'$ and its mean is $\mu_{i,x_j^*} = A_{ij}\, {\pmb x_{j}^*}$, and notice the covariance is not a function of the value of ${\pmb x_{j}^*}$.  Then by \citet{imhof_computing_1961} and \citet{scheffe_analysis_1959} one knows  for $Q_i = \sum_{k=1}^{d_i} x_{i,k}^{*2}$ we have
\begin{align}
Q_i \, |\, {\pmb x_{j}^*} \;  \sim \, \sum_{k=1}^{d_i} \lambda_k \chi_{1,k}^2(\delta_k^2)
\label{eqn:cond_imhof}
\end{align}
where the $\chi^2_{1,k}(\delta_k^2)$ are independent $\chi^2$ random variables on 1 degree of freedom and with non-centrality parameter $\delta_k^2$, for $\delta_k = v_k' \, \mu_{i,x_j^*} / \sqrt{\lambda_k}$  and for eigenvalues $\lambda_k$ and eigenvectors $v_k$ of $I-A_{ij}A_{ij}'$.




 Let $P(\;\cdot \;|\;\cdot \; )$ denote the conditional density of $Q_i$ given some event.  We can assert that 
\begin{align}
P(q_i \,| \, Q_j=l)=\int_{B_l}P( q_i \, |\, \pmb{x}_{dj}^*, Q_j=l)f(\pmb{x}_{d_j}^* \, | \, Q_j=l) \, d\,{\pmb x}_{d_j}^*= \int_{B_l}P(q_i \, |\, \pmb{x}_{d_j}^*)\, k  \, d\,{\pmb x}_{d_j}^*
\label{eqn:equalities}
\end{align}
The first equality 
holds because $P(q_i \,| \, Q_j=l) = \int_{B_l} P(q_i , \pmb{x}_{j}^* \, |\, Q_j=l) \, d\,{\pmb x}_{j}^* $ and

\noindent
$P(q_i , \pmb{x}_{j}^* \, |\, Q_j=l)  = P(q_i \, |\, \pmb{x}_{j}^*, Q_j=l)f(\pmb{x}_{j}^* \, | \, Q_j=l)$, where recall $B_l$ is the hypersphere of radius $\sqrt{l}$, the support of ${\pmb x}_{j}^*$ when conditioning on $Q_j=l$.

\vspace{0.1cm}

The second equality of (\ref{eqn:equalities}) holds because $f(\pmb{x}_{j}^* \, | \, Q_j=l)=k$, as written above, because $Q_j =\sum_{k=1}^{d_j} x_{j,k}^{*2}$ so values of ${\pmb x_{j}^{*}}$ consistent with the conditioning event are the hypersphere of radius $\sqrt{l}$.   Again, since $\pmb{x}_{j}^*$ has identity covariance and mean 0, that hypersphere follows a contour on the multivariate Gaussian density of constant value which we called $k$.  The second equality also assumes $\int_{B_l} P(q_i\, |\, \pmb{x}_{j}^*, Q_j=l) \, d\,{\pmb x}_{j}^*=\int_{B_l} P(q_i\, |\, \pmb{x}_{j}^*) \, d\,{\pmb x}_{j}^*$, which is true since $Q_i$ and $Q_j$ are conditionally independent given $\pmb{x}_{j}^*$.

So $P(q_i \,| \, Q_j=l)= \int_{B_l}P(q_i \, |\, \pmb{x}_{j}^*)\, k  \, d\,{\pmb x}_{j}^* $ based on (\ref{eqn:equalities}).  We know by (\ref{eqn:cond_imhof}) that $P(q_i \, |\, \pmb{x}_{j}^*)$ is the density of the sum of $d_i$ independent and scaled, non-central $\chi^2_1$ random variables with non-centrality and scaling parameters defined just below (\ref{eqn:cond_imhof}).

Above we introduced the density $f( x^*_{j,1} , \cdots ,  x^*_{j,d_j} \, | \, Q_j = l)$, from which we derived the constant density $k$ on $B_l$.  Now consider $f( x^*_{j,1} , \cdots ,  x^*_{j,d_j} \, | \, Q_j = m)$ for $l<m$, again a constant density on $B_m$, the surface of the hypersphere in ${\mathbb R}^{d_j}$ centered at 0 and of radius $\sqrt{m}$.
One way of arguing $Q_i$ is stochastically increasing in $Q_j$ relies on finding some bijection $g: B_l \rightarrow B_m $ under which these two densities are mapped between one another, and for which every element under $g$ is increasing absolutely, which will have implication on the non-centrality parameters of the $\chi^2_1$ random variables composing $Q_i$.  
In doing so, one can compare the mixture distributions arising in calculating $P(q_i | Q_j)$ for increasing $Q_j$.  We propose  $g({\pmb x} ) = \frac{\sqrt{m}}{\sqrt{l}}  {\pmb x}$ as one transformation satisfying these requirements.  Clearly $g$ is increasing absolutely in every element, and maps density $f( x^*_{j,1} , \cdots ,  x^*_{j,d_j} \, | \, Q_j = l)$ to $f( x^*_{j,1} , \cdots ,  x^*_{j,d_j} \, | \, Q_j = m)$, where the usual determinant of the Jacobian 
is $\Bigl(\frac{\sqrt{l}}{\sqrt{m}} \Bigr)^{d_j-1}$ rather than $\Bigl(\frac{\sqrt{l}}{\sqrt{m}} \Bigr)^{d_j}$ because of the quadratic constraint giving the spherical surface domain, and which intuitively is the ratio of the surface areas of $B_l$ and $B_m$.  That is, $f( x^*_{j,1} , \cdots ,  x^*_{j,d_j} \, | \, Q_j = m)= k\cdot r$ with $r=\Bigl(\frac{\sqrt{l}}{\sqrt{m}} \Bigr)^{d_j-1}$.  With this in mind, for some increasing $h(\cdot)$, we calculate expectations


\begin{align}
E[h(Q_i) | Q_j=l] = \int_0^\infty h(q_i) P(q_i|Q_j=l)  \, dq_i= \int_0^\infty \int_{B_l} h(q_i) P(q_i|{\pmb x_j^*})\, k  \, d {\pmb x_j^*}\, d q_i
\label{eqn:integral}
\end{align}
where the second equality holds because of (\ref{eqn:equalities}).  Then
\begin{align}
\int_0^\infty \int_{B_l} h(q_i) P(q_i|{\pmb x_j^*}) \, k \, d {\pmb x_j^*} \, d q_i = \int_{B_l}  \int_0^\infty h(q_i) & P(q_i|{\pmb x_j^*})\,  k \, d q_i\, d {\pmb x_j^*}  \label{eqn:integral}  \\
 <  \int_{B_l}  \int_0^\infty  h(q_i) P(q_i|g({\pmb x_j^*})) k  \, d q_i \, d {\pmb x_j^*}  &= \int_{B_m}  \int_0^\infty h(q_i) P(q_i|{\pmb x_j^*})\, k\cdot r \,  d q_i \, d {\pmb x_j^*} 
 \label{eqn:integral2}
\end{align}
where (\ref{eqn:integral}) holds by Fubini's theorem because $P(q_i | {\pmb x_j^*})$ is the density of a finite sum of scaled $\chi^2_1$ random variables with finite non-centrality parameters, and so has finite variance, and the measure of $B_l$ is finite.  The inequality in (\ref{eqn:integral2}) holds because $g({\pmb x})=\frac{\sqrt{m}}{\sqrt{l}}{\pmb x}$ is strictly increasing and $\chi^2_1$ random variables are strictly stochastically increasing in their non-centrality parameters, which here are 
$\delta_k^2 = (v_k' \, A_{ij} \, {\pmb x}_{j}^* / \sqrt{\lambda_k})^2$
for all $k=1 \dots d_i$ (ie, $\bigl(v_k' \, A_{ij} \, {\pmb x}_{j}^* / \sqrt{\lambda_k}\bigr)^2 < \bigl(v_k' \, A_{ij} \, g({\pmb x}_{j}^*) / \sqrt{\lambda_k}\bigr)^2$ for ${\pmb x}_{j}^* \neq 0$ for all $k$), and the equality in (\ref{eqn:integral2}) holds by change of variable.

But then the last expression in (\ref{eqn:integral2}) is $E[h(Q_i) | Q_j = m]$, giving $$E[h(Q_i) | Q_j = l]<E[h(Q_i) | Q_j = m]$$
as desired for $l<m$ and increasing $h(\cdot)$. So $Q_i$ is stochastically increasing in $Q_j$.  Since $i$ and $j$ are arbitrary and the result holds simultaneously on all other $i\neq j$, we have PRDS on $(Q_1 \dots Q_m)$.

\qed

\section{Consistency of the Proposed test statistics}

{\bf Proof of Lemma 1}

We suppress the stratum subscript $l$ notation in what follows in favor of $n$ to emphasize each quantity's dependence on sample size $n$ as it gets large.  In showing this lemma, we rely in part on results of \citet{golub_generalized_1979}, \citet{craven_smoothing_1979}, and \citet{reiss_smoothing_2009} for showing $\tilde{\lambda}_n = o(n)$ under either GCV or REML estimation.  With this fact and some manipulation of covariance matrices we then show the asymptotic $\chi_{d+1}^2$ null distribution of test statistics $T_k$ with the multivariate CLT and Slutzsky's theorem.  


Let $Z_n$ be the within stratum $n \times m$ design matrix with $n$ getting large.  A rotation and appropriate scaling of it will allow expression of $(Z_n^T Z_n + \lambda S)$ in the ridge regression shrinkage structure whose GCV estimation Golub et al characterize in Technometrics, 1979.  On page 219 of \citet{golub_generalized_1979}, the authors write that both $\lambda_0$, a minimizing sequence of a criterion related to GCV, and $\tilde{\lambda}$, a minimizing sequence of the GCV criterion, satisfy $n \lambda^{1/m} \rightarrow 0$ with $m>1$, the latter having been shown in a related problem in \citet{craven_smoothing_1979} and ultimately giving $\lambda = o(n)$ for GCV estimation.  One can likewise examine results in \citet{wahba_comparison_1985} for confirmation of \citet{golub_generalized_1979}.  

If estimating ${\lambda}$ with REML, one can examine \citet{reiss_smoothing_2009} and see that a manipulation of their equation (8) yields 

\begin{equation}
2\lambda  = \frac{(n-p)}{\mbox{tr}(P_\lambda)} \cdot \frac{y^T P_\lambda^2 y}{y^T P_\lambda y }
\label{eqn:ogden}
\end{equation}

One can then use their simplifications of the quadratic form and trace terms from their equation (11) and equation (12) and see that for all $\lambda$ the numerator and denominator quadratic forms $y^T P^k_\lambda y$, $k=1,2$, in the RHS are $O(n)$ and have quotient $O(1)$.  Likewise, the numerator and denominator of the $(n-p)/\mbox{tr}(P_\lambda)$ term are $O(n)$ and have quotient $O(1)$.  Thus the only sequence of $\lambda$ for which equality in equation (\ref{eqn:ogden}) holds is $O(1)$, giving $\lambda = o(n)$ under REML.  Fittingly, it should be our expectation that either REML or GCV, two commonly used fitting criteria in $L_2$ penalized settings, must be $o(n)$ for penalization-induced bias to converge to 0. which it would do asymptotically for any well performing fitting criterion.  

This in mind, we consider the proposed test statistics $T_k$ asymptotically for the sequence $\{\tilde{\lambda}_n \}$ when $\tilde{\lambda}_n = o(n)$.  First factorize full rank $Z_n$ of dimension $n\times m$ into $Z_n = Q_n R_n$ with upper triangular and square matrix $R_n$ being all non-negative elements, which is unique, and $Q_n^T Q_n = nI$, this $Q_n$ `absorbing' the $n$ term, with $I$ identity of dimension $m$.  Likewise set

$$\tilde{S}_n =\tilde{\lambda}_n S $$ where  $\tilde{\lambda}_n$ is estimated under REML or GCV.

Assuming that for row vector $Z_i$ of matrix $Z_n$ we have $Var(Z_i) = \Sigma_Z = R^T R$ for all $1\leq i\leq n$, $R_n \rightarrow R$ because of uniqueness of the factorization. As established above, $\tilde{\lambda}_n = o(n)$ under GCV or REML fitting. We can factorize $R_n^{-1}\tilde{S_n} R_n^{-T}=\tilde{\lambda}_n \, U_n^T D_n U_n$, where $D_n$ is diagonal and may have zeros along it, while $U_n$ is orthonormal with $U_n \rightarrow U$ and $D_n \rightarrow D$ uniformly for finite-dimensional $U_n$ and $D_n$.

Using these decompositions and as a variation on the simplification of \citet{marra_coverage_2012}, we have the following expression:
\begin{align}
(Z_n^T Z_n + \tilde{S}_n)^{-1} =& (R_n^T nI R_n + R_n^T (R_n^{-T} \tilde{S}_n R_n^{-1}) R_n)^{-1} \nonumber  \\ 
=& \;(R_n^T (nI + \tilde{\lambda}_n \, U_n^T D_nU_n) R_n)^{-1} \nonumber  \\ 
=& \;(R_n^T U_n^T (nI + \tilde{\lambda}_n \, D_n) U_n R_n)^{-1} \nonumber \\ 
=& \;(U_n R_n)^{-1} (nI + \tilde{\lambda}_n \, D_n)^{-1} (U_n R_n)^{-T}
\label{eqn:simp}
\end{align}
We can use (\ref{eqn:simp}) then to simplify
\begin{align*}
\widehat{\mbox{Var}}(\hat{{\pmb b}}_n) &= \widehat{\mbox{Var}}\big((Z_n^T Z_n + \tilde{S}_n)^{-1} Z_n^T {\pmb y}_n\big) \\ 
&= (Z_n^T Z_n + \tilde{S}_n)^{-1} Z_n^T \, \widehat{\mbox{Var}}({\pmb y}_n) \, Z_n (Z_n^T Z_n + \tilde{S}_n)^{-1} \\
&= \hat{\phi} \, (U_n R_n)^{-1} (nI + \tilde{\lambda}_n \, D_n)^{-1} (U_n R_n)^{-T}  Z_n^T Z_n
(U_n R_n)^{-1} (nI + \tilde{\lambda}_n \, D_n)^{-1} (U_n R_n)^{-T} \\ 
&= \hat{\phi} \, (U_n R_n)^{-1} (nI + \tilde{\lambda}_n \, D_n)^{-1} U_n^{-T}R_n^{-T}  R_n^T Q_n^T Q_n \, R_n
R_n^{-1} U_n^{-1} (nI + \tilde{\lambda}_n \, D_n)^{-1} (U_n R_n)^{-T} \\ 
&= \hat{\phi} \, (U_n R_n)^{-1} (nI + \tilde{\lambda}_n \, D_n)^{-1} nI (nI +\tilde{\lambda}_n \,  D_n)^{-1} (U_n R_n)^{-T}
\end{align*}
Pulling out $n$ from the inverses and collecting terms yields
$$\widehat{\mbox{Var}}(\hat{{\pmb b}}_n) =\hat{\phi} \, (U_n R_n)^{-1} (I + \tilde{\lambda}_n/n \, D_n)^{-1} I/n (I +\tilde{\lambda}_n/n \,  D_n)^{-1} (U_n R_n)^{-T}$$
Since $\tilde{\lambda}_n = o(n)$, $\tilde{\lambda}_n/n \rightarrow 0$, the expression converges to
\begin{align}
\hat{\phi} \,  R_n^{-1}  U_n^{-1}  \,I/n \,U_n^{-T}  R_n^{-T} =\hat{\phi} \, R^{-1}_n R^{-T}_n \, /n 
\label{eqn:var}
\end{align}
Likewise the covariance estimator from the Bayesian perspective yields the same expression asymptotically
\begin{align*}
\hat{\phi} (U_n R_n)^{-1} &(nI +\tilde{\lambda}_n \, D_n)^{-1} (U_n R_n)^{-T} \\ 
& = \hat{\phi}/n \,(U_n R_n)^{-1}  (I +\tilde{\lambda}_n/n \, D_n)^{-1} (U_n R_n)^{-T} 
\end{align*}
Again since $\tilde{\lambda}_n = o(n)$, $\tilde{\lambda}_n/n \rightarrow 0$, as with (\ref{eqn:var}) the expression converges to
\begin{align}
& \hat{\phi} \, R^{-1}_n R^{-T}_n \, /n 
\label{eqn:bayes}
\end{align}
Considering $\hat{{\pmb b}}_n$ and using again the simplification in (\ref{eqn:simp}) we have
\begin{align*}
\hat{{\pmb b}}_n =& (Z_n^T Z_n + \lambda_n S)^{-1}Z_n^T{\pmb y}_n \\ 
= &((U_n R_n)^T (nI + \tilde{\lambda}_n  D_n )  U_n R_n )^{-1} R_n^T Q_n^T {\pmb y}_n \\
= &(U_n R_n)^{-1} (nI + \tilde{\lambda}_n  D_n )^{-1}  U_n^{-T} R_n^{-T} R_n^T Q_n^T {\pmb y}_n 
\end{align*}
which converges to
\begin{align*}
= & R_n^{-1} U_n^{-1} I/n  U_n^{-T} Q_n^T {\pmb y}_n = R_n^{-1} Q_n^T {\pmb y}_n/n
\end{align*}
\noindent
a quantity we note is $O(1)$ because $Q_n^T Q_n = nI$.

Assuming without loss of generality $E(\hat{\pmb{b}}_n)={\pmb b_0} + o(1)$, then multiplying $\hat{{\pmb b}}_n-{\pmb b_0}$ by the inverse root of expression (\ref{eqn:var}) yields
\begin{align*}
 (n/\hat{\phi})^{1/2}  R_n ( \hat{{\pmb b}}_n -{\pmb b_0})= &(n/\hat{\phi})^{1/2} R_n  ( R_n^{-1} Q_n^T {\pmb y}_n/n -{\pmb b_0}) \\
 \hspace{2cm}= &(n/ \hat{\phi})^{1/2}  (Q_n^T {\pmb y}_n/n - R_n {\pmb b_0})  \sim MVN(0, I)
\end{align*}
by the multivariate central limit theorem 
and where we have used Slutzsky's theorem in using the quantities to which we've converged. 

Since the $\hat{\pmb{b}}_l$ with $l \in \{1,2 \}$ used in the test statistics $T_k$ are fit on independent strata, the variance of their difference is additive, and it follows that constructing $T_k$ with $(d+1)$-length subvectors of $\hat{\pmb{b}}_1 -\hat{\pmb{b}}_2$ under the null hypothesis $H^0_k$ of these subvectors' equivalence, and corresponding submatrices of $V_1 + V_2$, yields a $\chi^2_{d+1}$ null distribution asymptotically.  
\qed

\section{Inverting multidiagonal Toeplitz matrices}
\noindent We leverage an analytic result on the inverse of Toeplitz matrices to argue that the PRDS condition holds.  To do so, we show that the covariance of distant (along the support of the smooth) quadratic form test statistics goes to zero quickly, and that of proximal quadratic forms is positive.  We then rely on conventional thinking that for broad families of especially unimodal distributions, positive correlation is sufficient for PRDS.  

We assume that the design points modelled with a smooth are distributed uniformly over the covariate's support and aligned with uniformly distributed knots so that the information matrix ${\pmb I_T}$ is Toeplitz with small modification to the two corner elements of the diagonal.  One can alternatively assume that, regardless of alignment with knots, design points are uniformly distributed and sufficiently dense so that deviations from Toeplitz structures are bounded by an arbitrary $\epsilon>0$ uniformly for each element in the matrix.  One can also assume that if at the boundary, values deviate slightly from a Toeplitz structure, one can modify the placement of knots or shape of basis functions so that the structure is achieved.

\begin{proof}[{\bf Proof of Lemma \ref{factor}}]
Following \citet{wang_explicit_2015} and \citet{montaner_five-diagonal_1995}, then it is apparent that $2\lambda$ is a root of the polynomial $f_1(s) = s^3 - \epsilon s^2 + (\theta^2 - 4 \lambda^2) s + (2\theta^2 \lambda - 4 \epsilon \lambda^2)$, giving $\zeta_1 + \zeta_2 = 2\lambda$. Again following \citet{wang_explicit_2015,montaner_five-diagonal_1995}, the roots of $f_2(s) = s^2 - 2\lambda s + \lambda^2$ then give $\zeta_1,\zeta_2$ which are found to be $\lambda$.  Let the roots of $f_3(s) = s^2 - \lambda s + \lambda(\epsilon - 2\lambda)$ be $\pi_1$ and $\pi_2$, to be used in the factorization.  Then $\pi_1, \pi_2$ are $\theta/2 \pm \sqrt{\theta^2 - 4\lambda(\epsilon - 2\lambda)}/2$, which are real for $\theta^2 - 4\lambda(\epsilon - 2\lambda)\geq 0$.  So $Z_1$, $Z_2$ are of the same dimension as $P$ and tridiagonal toeplitz, with elements in order along the diagonal $(\lambda,\pi_1,\lambda)$ and $(1,\pi_2/\lambda,1)$ for each matrix respectively.  
\end{proof}

\begin{proof}[{\bf Proof of Theorem \ref{log_linear}}]
There exist 2 unique, real roots of the polynomial $f(s) = \lambda s^2 + \pi_i s + \lambda$, $i=1,2$, provided $\pi_i^2 > 4 \lambda^2$.  We can therefore simplify \citet{dow_explicit_2003} to elements of $Z_i^{-1}$ defined with
$$z_{kl}^{(i)} = z_{lk}^{(i)} = \frac{(-1)^{l-k} \sinh (\psi_i k ) \cdot \sinh(\psi_i (n+1-l))}{\sinh \psi_i \cdot \sinh (\psi_i(n+1))}$$
for $k\leq l$, $\psi_i = \operatorname{arcosh} (\pi_i/(2\lambda))>0$ by assumption for positive $\pi_i$ and $\lambda$.  It is the case that $Z_1^{-1} =1/\lambda \cdot \bigl(z_{kl}^{(1)} \bigr)$ and  $Z_2^{-1} = \bigl(z_{kl}^{(2)} \bigr)$.  Because $\sinh x / \exp{x} \rightarrow 1/2$ quickly for increasing $x$ and $\exp x> 2 \sinh x$ for $x>0$ , we can well-approximate and upper bound in absolute value with $z_{lk}^{(i)} \approx (-1^{l-k}\cdot K^{(i)}) \exp  (\psi_i (k-l))$ for indices $k$ getting large and $l$ getting small and a constant 
$K^{(i)}=\bigl(\exp (\psi_i(n+1))\bigr)/\bigl(4\, \sinh \psi_i \cdot \sinh (\psi_i(n+1))\bigr)$.  As the dimension of $P$ gets large, the proportion of elements converging to this quantity goes to 1.  Now consider element $r,t$ of $P^{-1}=  Z_2^{-1} Z_1^{-1}$ with $r\leq t$ because $P^{-1}$ is symmetric, which is calculated 
\begin{equation}
\sum_{i=1}^{m_T+d-1}  z_{ri}^{(2)} z_{it}^{(1)} \approx (-1^{t-r}\cdot K^{(1)} \, K^{(2)})  \sum_{i=1}^{m_T+d-1} \exp{(-\psi_2 |i-r| - \psi_1 |t-i|)}
\label{approx}
\end{equation}
Assuming without loss of generality that $\psi_2 \leq \psi_1$ the sum can be partitioned into three finite series by considering collections of those terms whose sum of indices increase, decrease, or are constant (for which one considers $\psi_2 <\psi_1$ and $\psi_2 =\psi_1$ separately), which can be shown to be respectively
\begin{align*}
\Gamma&= \exp{(-(t-r) \psi_2)}\frac{\exp{(-|\psi_1+\psi_2|)}-\exp{(-r\, |\psi_1 + \psi_2|)}}{1-\exp{(-|\psi_1+\psi_2|)}}  \\
\Lambda &= \exp{(-(t-r) \psi_1)}\frac{\exp{(-|\psi_1+\psi_2|)}-\exp{(-(n-t+1) |\psi_1 + \psi_2|))}}{1-\exp{(-|\psi_1+\psi_2|)}} \\
\Delta_{\ne} &= \exp{(-(t-r) \psi_2)}\frac{1-\exp{(-(t-r+1) |\psi_1 - \psi_2|)}}{1-\exp{(-|\psi_1-\psi_2|)}}  \\
\Delta_{=} &=   \exp{(-(t-r)\psi_2)}\, (t-r + 1) 
\end{align*}
where $\Delta_{\ne}$ is the relevant term for the case where $\psi_2<\psi_1$ and $\Delta_=$ otherwise. For $r$ increasing and $t$ decreasing, which occur simultaneously for increasing dimension of $P$, the sum scaled by the appropriate sign then converges to  
\begin{equation*}
\begin{split}
 (-1^{t-r} K^{(1)}K^{(2)}/ \lambda)(\Gamma + \Delta_{*} + \Lambda ) \rightarrow (P^{-1})_{r,t}
\end{split}
\end{equation*}
for $\Delta_*$ as $\Delta_=$ or $\Delta_{\ne}$ according to the case.  For increasing $(t-r)$, the two terms $\Gamma$ and $\Delta_{\ne}$ or single term $\Delta_=$ dominate the sum for $\psi_2 < \psi_1$ or $\psi_2 = \psi_1$, respectively.  In the former case it is evident that because the fractions in $\Gamma$ and $\Delta_{\ne}$ converge to constants, a $1$ unit increase in $t$ or decrease in $r$ scales the sum by $\exp{(-\psi_2)}$.  For $\psi_2 = \psi_1$, the $\Delta_=$ term associated with a 1 unit movement away from the diagonal vertically or horizontally likewise converges to a proportional change of $\exp{(-\psi_2)}$. Lastly because equation (\ref{approx}) is an upper bound in absolute value, the rate of decay $\min_i \psi_i$ is at least preserved in the elements of symmetric $P^{-1}$ for indices $r$ small and $t$ large, $r\leq t$.
\end{proof}

\section{Additional simulations}

\newgeometry{left=0.75cm,right=1.4cm}


\begin{figure}[H]
\begin{center}
\includegraphics[width=0.99\textwidth]{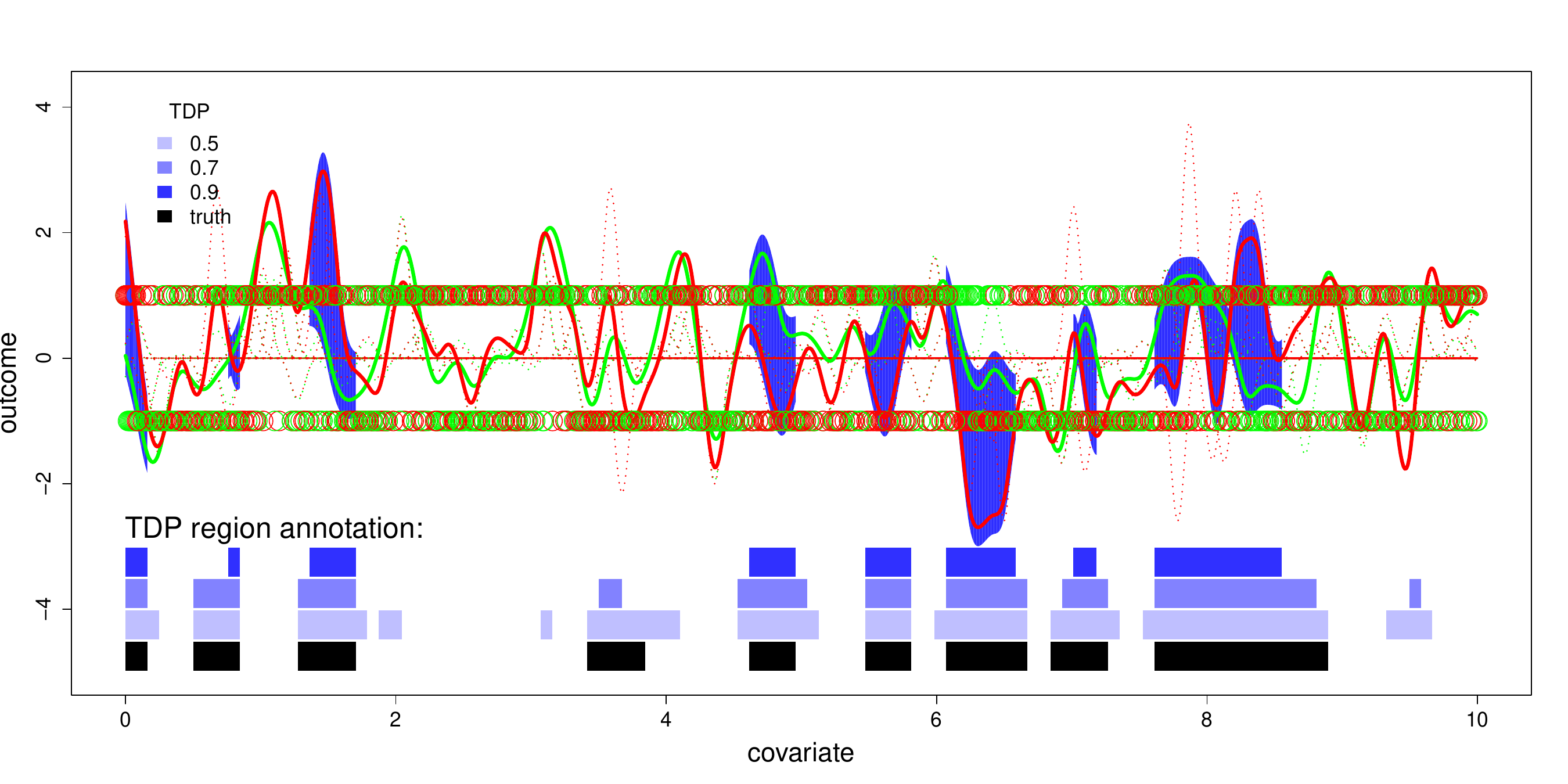}
\captionsetup{width=0.98\textwidth}
\caption{Two simulated curves for a binary outcome on a domain of (0,4.5) with highlighted difference regions along the smooths in dark blue.  The data are fit using logistic regression and the smooths shown correspond to the estimated linear predictors.  The points plotted along 1 and -1 of the y-axis are the modelled 1's and 0's, respectively, colored according to the corresponding smooth, and drawn on the graph in a random order so that their shade of color communicates the relative quantity of each.  The highlighted dark blue region corresponds to the 0.9 TDP annotation of the same color shown at the bottom of the figure.  There is analogous `bar' annotation for estimated TDP's of 0.7 and 0.5 in different shades of blue.  The black region at the most bottom shows the intervals where the 2 curves are generated from different basis functions.  The many dotted line curves in red and orange show the underlying basis functions scaled according to the true basis coefficients.  The estimates of their superimpositions -- the estimated linear predictor smooths -- are the thicker, solid red and orange curves.  We see that the TDP region annotation of 0.5, 0.7, and 0.9, are relatively accurate estimates of TDP as compared to the truth, also confirmed in Figure \ref{asymp_tdp_binary}.  This figure corresponds to a minimum effect size delta of 3.38 in the difference regions.}
\label{eff_338}
\end{center}
\end{figure}

\begin{figure}[H]
\centering
\begin{subfigure}{0.31\textwidth}
\includegraphics[width=\linewidth]{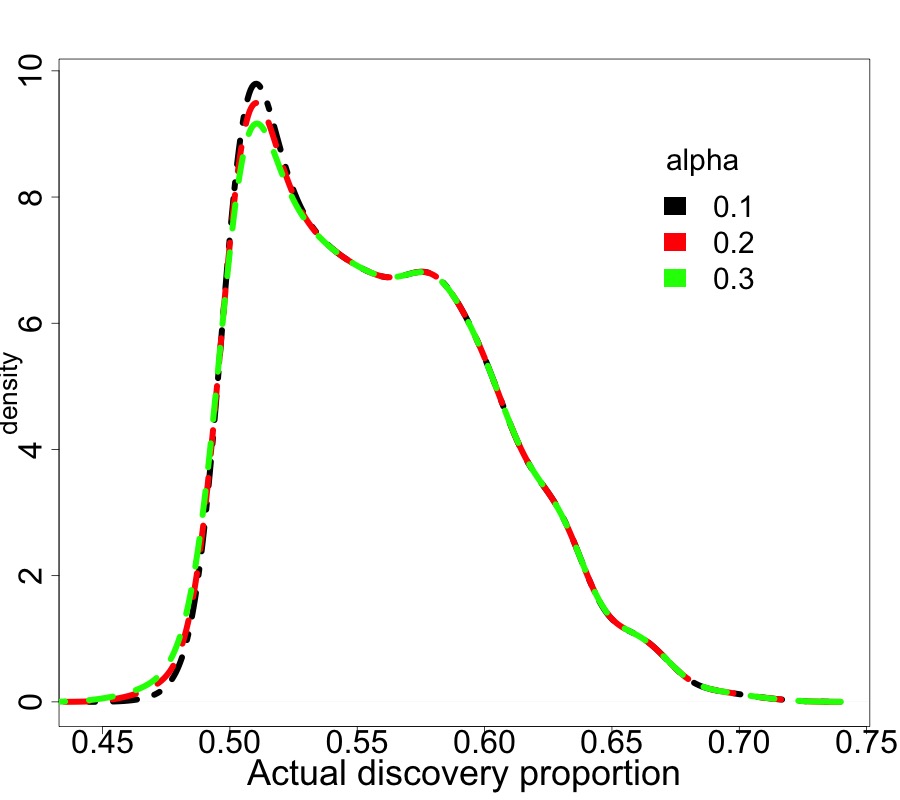}
\caption{Actual TDP for an estimate of 0.5} 
\label{tdp_dist1_30}
\end{subfigure}
\begin{subfigure}{0.31\textwidth}
\includegraphics[width=\linewidth]{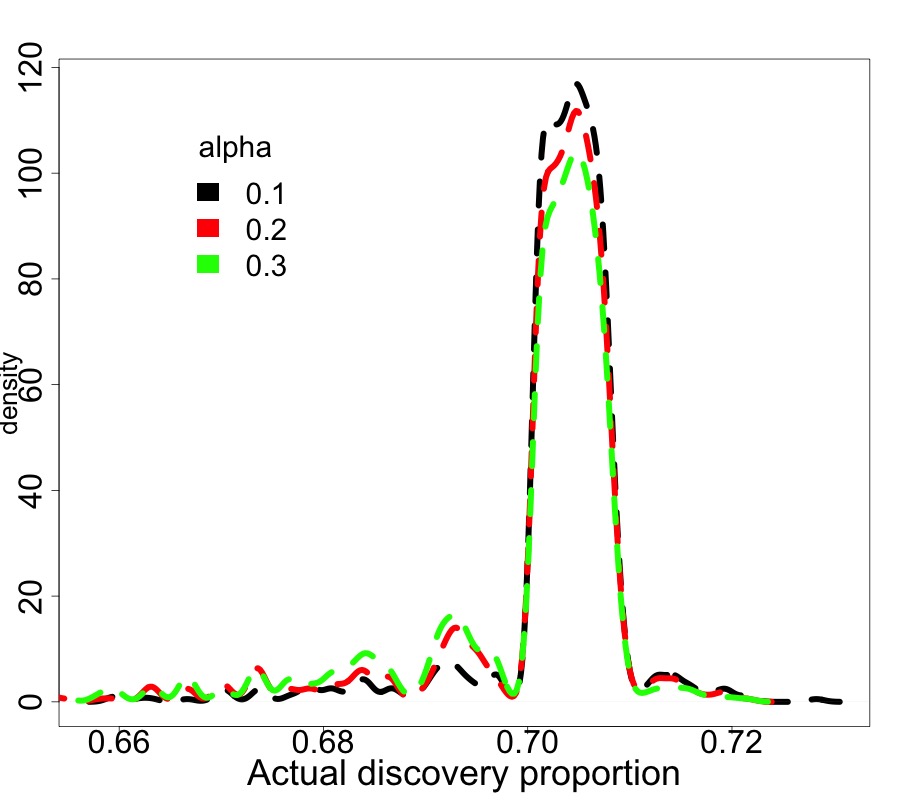}
\caption{Actual TDP for an estimate of 0.7} 
\label{tdp_dist2_30}
\end{subfigure}
\begin{subfigure}{0.31\textwidth}
\includegraphics[width=\linewidth]{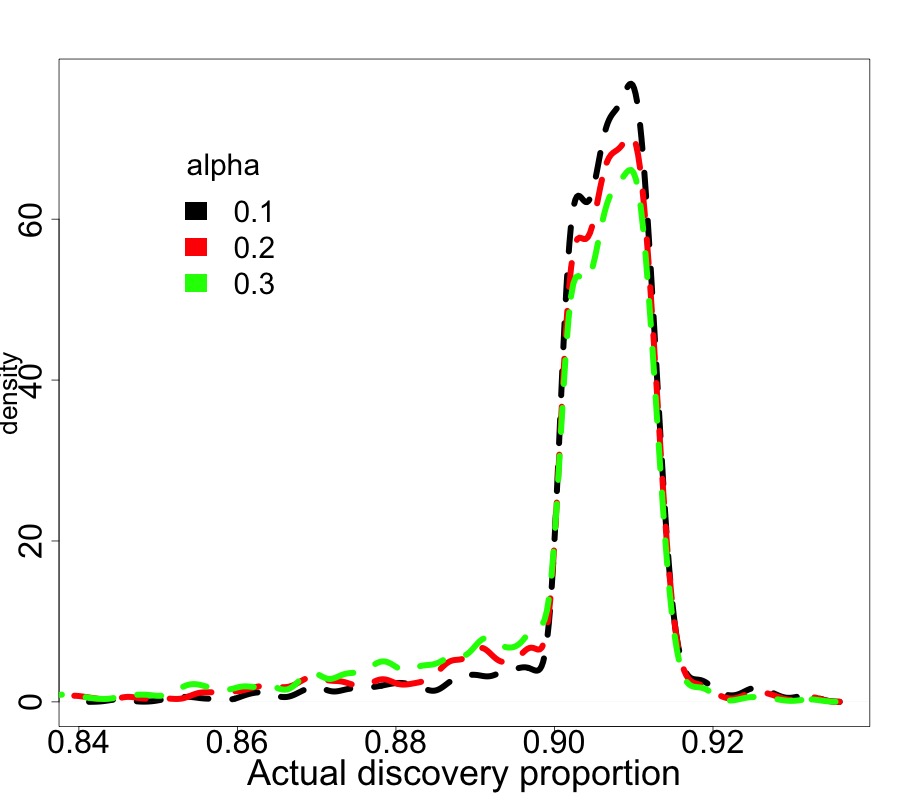}
\caption{Actual TDP for an estimate of 0.9} 
\label{tdp_dist3_30}
\end{subfigure}
\captionsetup{width=0.8\textwidth}
\caption{Distribution of actual TDP over 1000 iterations with 30 non-zero $\Delta b_k$'s for estimates of a) 0.5, b) 0.7, and c) 0.9, under different $\alpha$'s, specified in the legend.  Distributions were estimated using a kernel density estimator.  One observes all distributions to be relatively narrow, though less so with the estimated TDP of 0.5.  One sees slightly lower density around each distribution's mode for increasing $\alpha$, and slightly higher density around the tails.} 
\label{tdp_dist30}
\end{figure}

\begin{figure}[H]
\begin{center}
\includegraphics[width=0.5\textwidth]{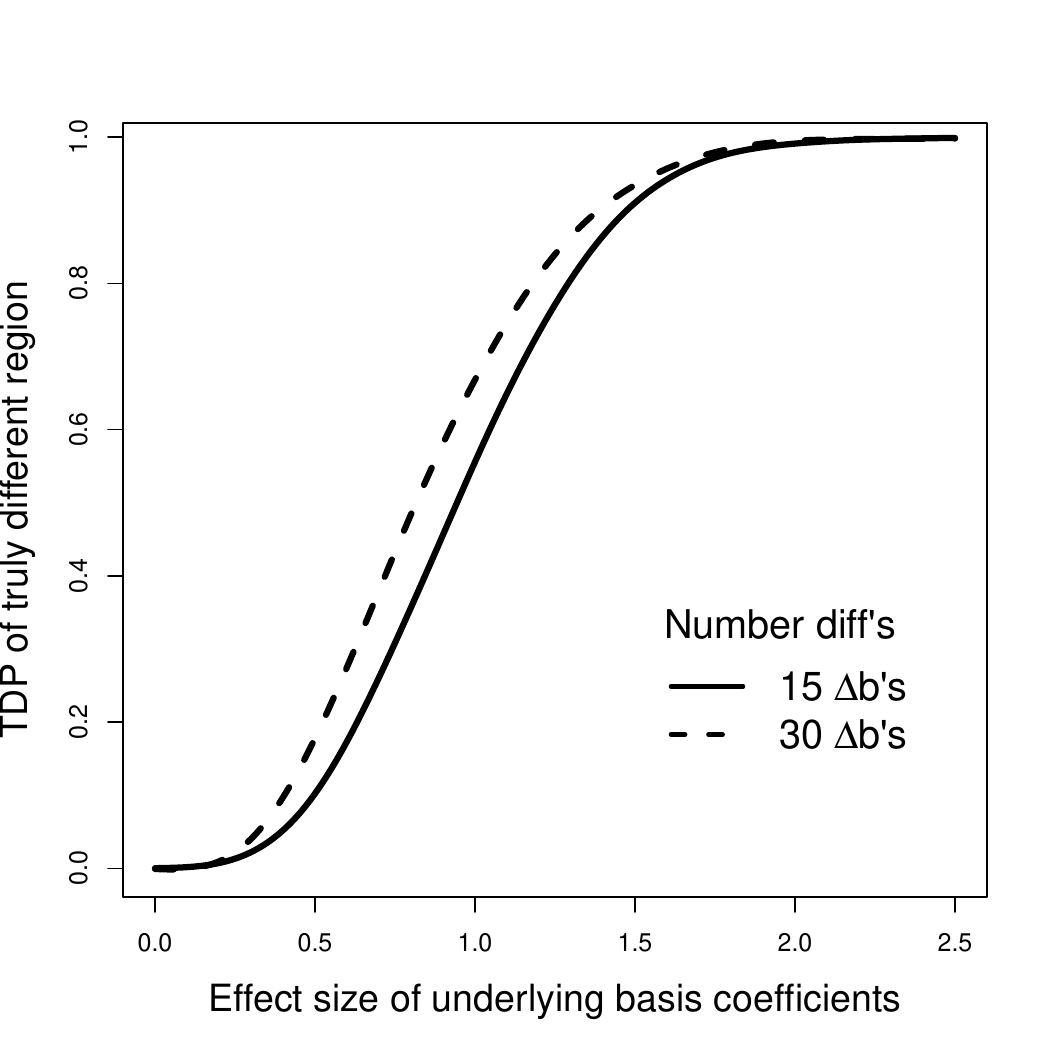}
\captionsetup{width=0.7\textwidth}
\caption{Estimated TDP calculated over a range of effect sizes for regions that are truly different, with an $\alpha$ of 0.2.  Those truly different region are composed of 15 or 30 non-zero $\Delta b_k$'s, many of which are contiguous, among a background of 105 or 90 zero $\Delta b_k$'s, respectively, totalling 120 $\Delta  b_k$'s. The lines in the figure are smooth curves fit over the underlying calculated TDP's. } 
\label{tdp_tru_diff}
\end{center}
\end{figure}

\begin{figure}[H]
\centering
\begin{subfigure}{0.31\textwidth}
\includegraphics[width=0.99\linewidth]{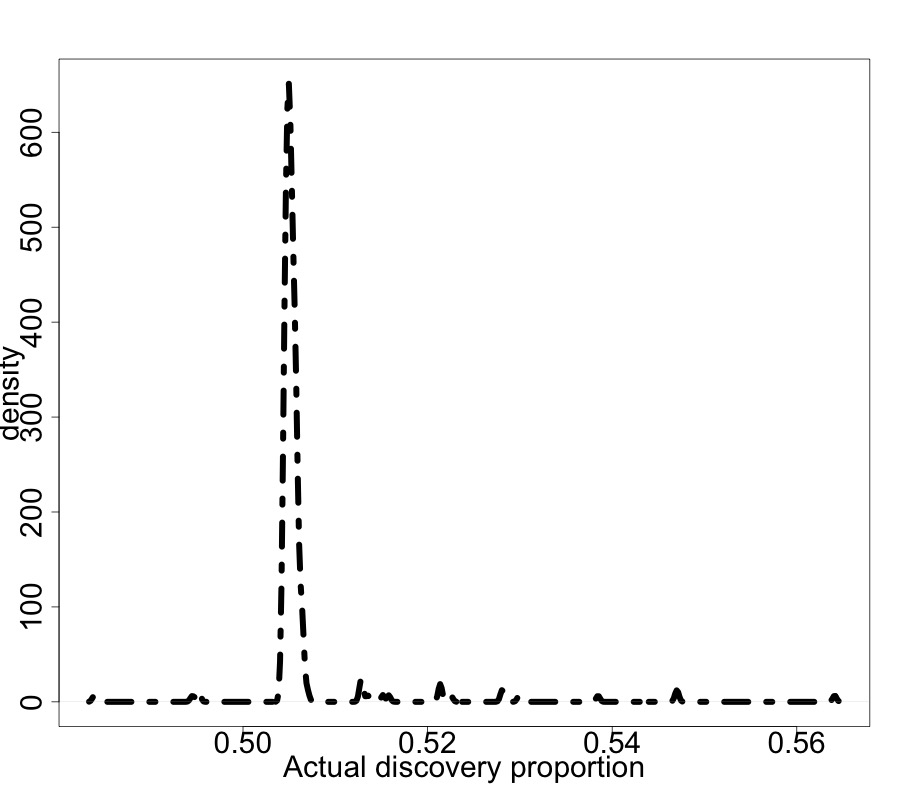} 
\caption{Empirical discovery proportion for nominal rate of 0.5.}
\label{label:} 
\end{subfigure}
\begin{subfigure}{0.31\textwidth}
\includegraphics[width=0.99\linewidth]{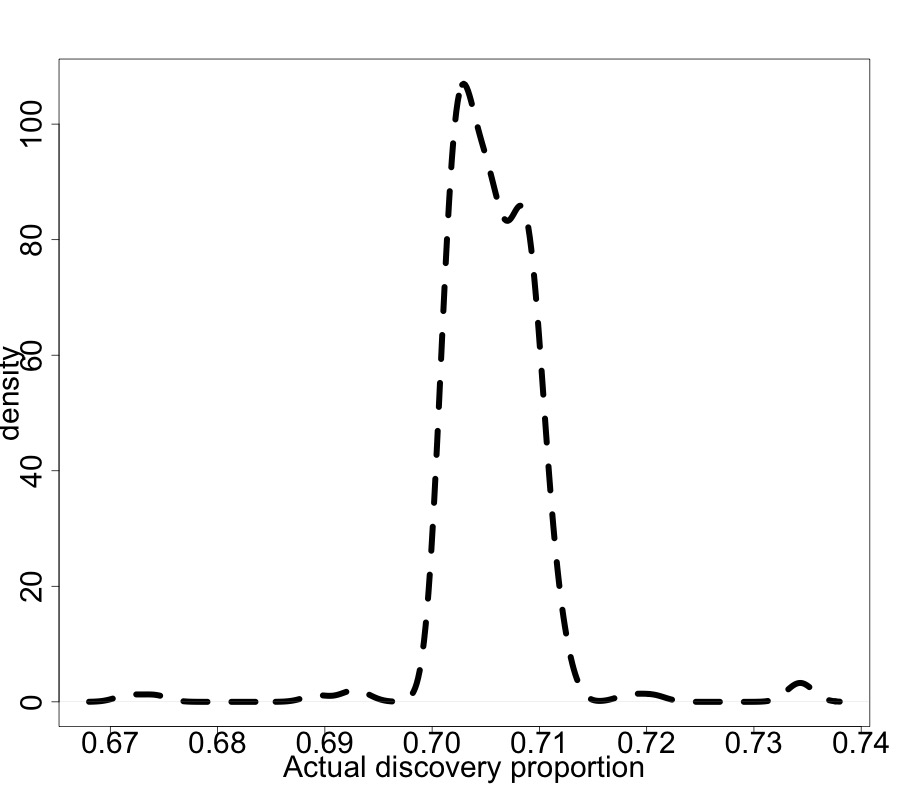} 
\caption{Empirical discovery proportion for nominal rate of 0.7.}
\label{label:} 
\end{subfigure}
\begin{subfigure}{0.31\textwidth}
\includegraphics[width=0.99\linewidth]{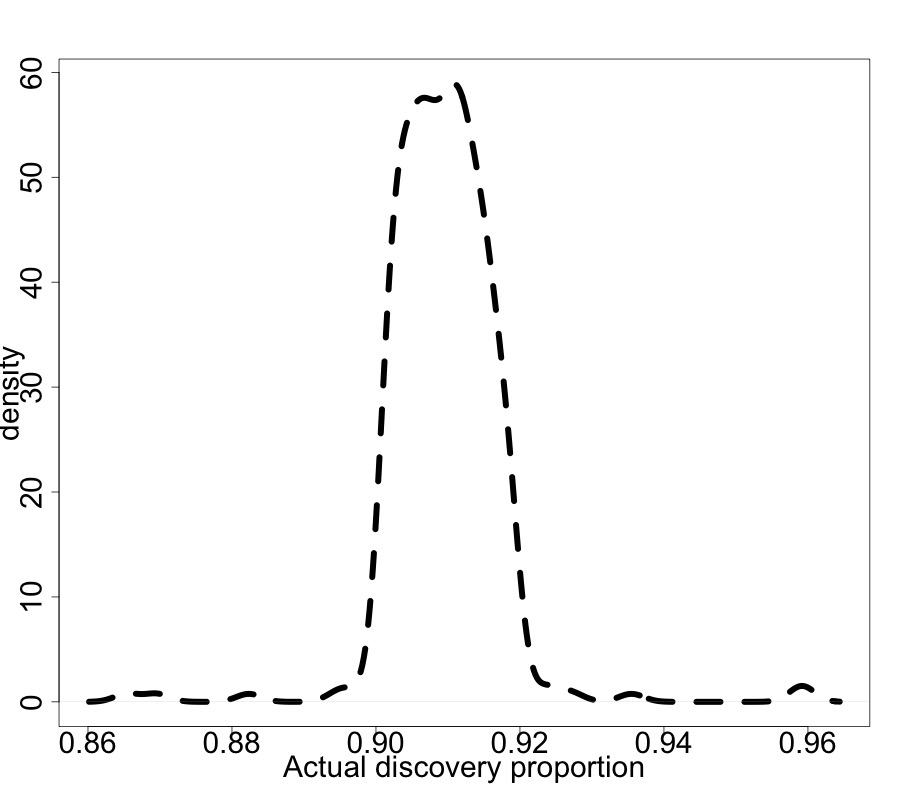} 
\caption{Empirical discovery proportion for nominal rate of 0.9.}
\label{label:} 
\end{subfigure}
\captionsetup{width=0.8\textwidth}
\caption{Empirical discovery proportion with 3 strata (smooths) for nominal rates of 0.5, 0.7, and 0.9 on an $\alpha=0.1$ when 25\% of the smooth is truly different.}
\label{fig:3strat}
\end{figure}





\restoregeometry





\end{document}